\documentclass[pre,floatfix,twocolumn,aps,showpacs,supergroupedaddress,epsfig,amsmath,amssymb]{revtex4}
\usepackage{epsfig,amsmath,amssymb,bm,epsf,graphics,psfrag,verbatim,subfigure}
\usepackage[usenames,dvipsnames]{color}
\usepackage{bbm}
\usepackage{mathrsfs}
\usepackage{amsfonts}
\usepackage{color}
\usepackage{pbox}
\usepackage{amssymb}
\usepackage{amsmath,amsthm}
\usepackage{graphicx}
\usepackage{dcolumn}
\usepackage{bm}
\usepackage{color,epsfig,multirow}
\usepackage{CJK}
\usepackage{subfigure}
\usepackage{verbatim}
\newcommand{\xv}{{\mathbf x}}
\newcommand{\yv}{{\mathbf y}}

\newcommand{\Yv}{{\mathbf Y}}
\newcommand{\rv}{{\mathbf r}}
\newcommand{\lv}{{\mathbf l}}

\newcommand{\Tr}{{\rm Tr}}

\newcommand{\be}{\begin{equation}}
\newcommand{\ee}{\end{equation}}
\newcommand{\ba}{\begin{eqnarray}}
\newcommand{\ea}{\end{eqnarray}}

\begin{document}

\title{A Multi-scale Monte Carlo Method for Electrolytes}
\date{\today}
\author{Yihao Liang$^{1}$}
\author{Zhenli Xu$^{2,3}$}
\author{Xiangjun Xing$^{1,2,4}$}
%\email{Correspondence author, email: xxing@sjtu.edu.cn}
%\footnote{Corresponding author, xxing@sjtu.edu.cn}}
\affiliation{$^1$Department of Physics and Astronomy, Shanghai Jiao Tong University, Shanghai, 200240, China \\
$^2$Institute of Natural Sciences, Shanghai Jiao Tong University, Shanghai, 200240, China \\
$^3$Department of Mathematics, and MOE Key Lab of Scientific and Engineering Computing,  Shanghai Jiao Tong University, Shanghai, 200240, China \\
$^4$Collaborative Innovation Center of Advanced Microstructures, Nanjing 210093, China}
%\author{Yihao Liang, Zhenli Xu, Xiangjun Xing}
% \affiliation{Department of Physics and Institute of Natural Sciences, Shanghai Jiao Tong University, Shanghai, China}

\begin{abstract}
Artifacts arise in the simulations of electrolytes using periodic boundary conditions (PBC).  We show the origin of these artifacts are the periodic image charges and the constraint of charge neutrality inside the simulation box, both of which are unphysical from the view point of real systems.  To cure these problems, we introduce a multi-scale Monte Carlo method, where ions inside a spherical cavity are simulated explicitly, whilst ions outside are treated implicitly using continuum theory.  Using the method of Debye charging, we explicitly derive the effective interactions between ions inside the cavity, arising due to the fluctuations of ions outside.  We find that these effective interactions consist of two types: 1) a constant {\em cavity potential} due to the asymmetry of the electrolyte, and 2) a {\em reaction potential} that depends on the positions of {all ions inside}.  Combining the Grand Canonical Monte Carlo (GCMC) with a recently developed fast algorithm based of image charge method, we perform a multi-scale Monte Carlo simulation of symmetric electrolytes, and compare it with other simulation methods, including PBC+GCMC method, as well as large scale Monte Carlo simulation.  We demonstrate that our multi-scale MC method is capable of capturing the correct physics of a large system using a small scale simulation.  

%By contrast, we show that the popular MC method based on periodic boundary condition and Ewald summation suffers from artifacts caused by artificial periodic image charges.

%We will present simulation results for the homogeneous electrolyte and show the importance of the total charge's fluctuation.

%{\bf Keywords:} Electrostatics, Monte Carlo, Charged systems, Electrolytes, Ewald Summation, Multi-scale.
\end{abstract}
% 52.25.Kn
\maketitle
\section{Introduction}

Monte Carlo (MC) simulation of Coulomb many body systems is a difficult problem \cite{Valleau-1980,frenkel2002book}, mainly due to  the long range nature of Coulomb interaction.  In the most na\"ive simulation strategy, one would confine a collection of ions inside a box with hard walls, and compute the total energy by adding up all pairwise interactions. This leads to two obvious difficulties: 1) The artifacts of hard wall propagate into the bulk of the system, with the characteristic scale set by the Debye length.   This renders large amount of simulation data useless.  The situation becomes worse in the dilute limit.  2) The computational complexity of each Monte Carlo cycle is of order of $N^2$, with $N$ the system size.  This makes simulation of large size systems practically impossible.  Obviously, both difficulties have their roots in the long range nature of Coulomb interaction.

Adoption of periodic boundary conditions (PBC) restores the translational symmetry and hence eliminates all the boundary effects.  Furthermore, one can take advantage of periodicity and compute the long range part of Coulomb energy in Fourier space, using the method of Ewald summation  \cite{Ewald:AP:1921}.  (For a pedagogical introduction, see \cite{frenkel2002book}.) This reduces the computational complexity from $N^2$ to $N^{3/2}$, which is still not fast enough for large scale simulations.  For molecular dynamics (MD) simulations, one can use particle-mesh method and fast Fourier transform \cite{DYP:JCP:1993,HE:1988} to reduce to complexity further down to $N \log N$.  Unfortunately, these techniques are not applicable for Monte Carlo simulations.  This is probably the main reason why MD has become the more popular method for charged systems, even when one is only insterested in static properties. 

\begin{figure*}[th!]
\subfigure[]{
\includegraphics[width=0.4\textwidth]{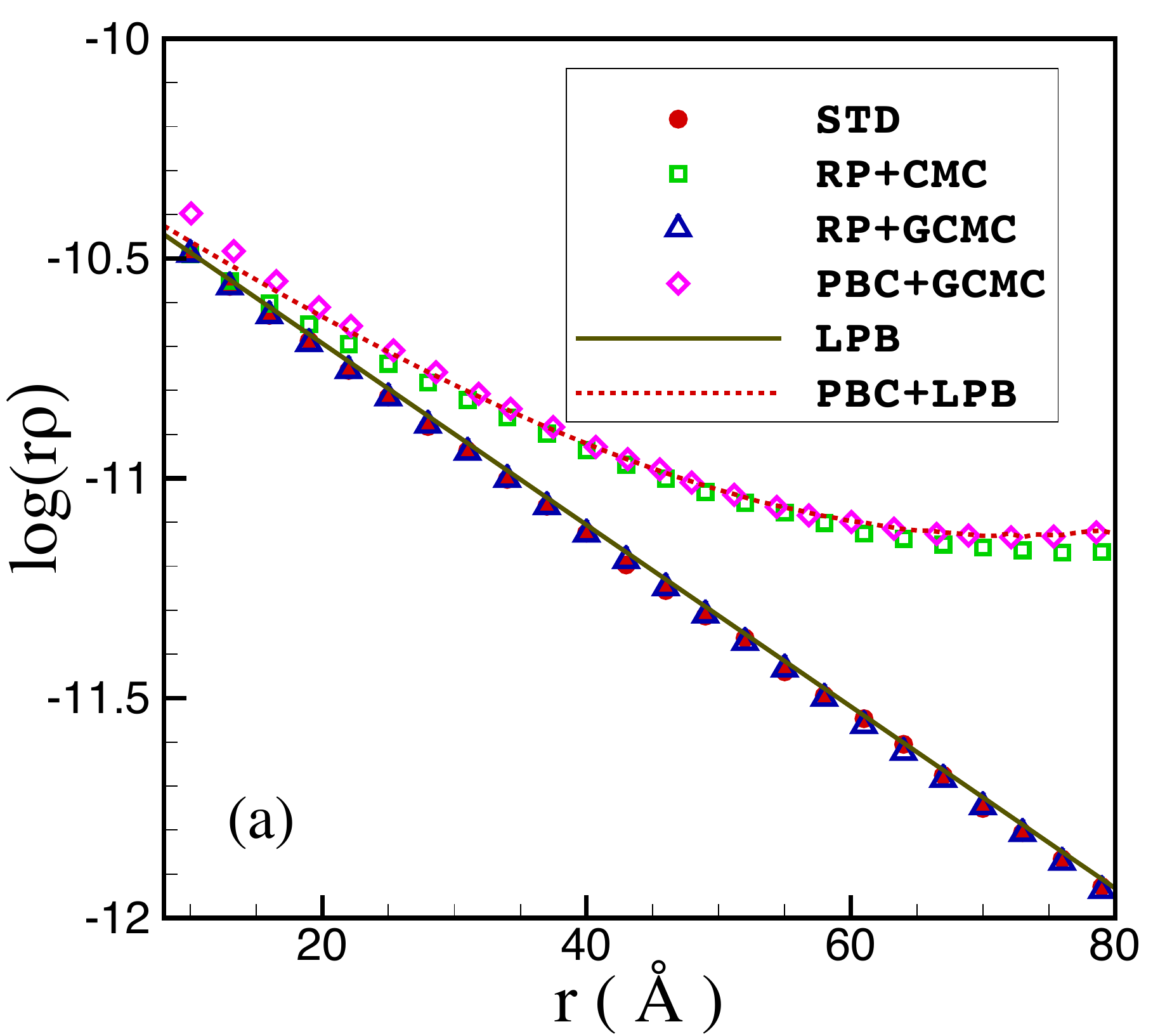}
\label{Fig_screening-rho}
}
\hspace{15mm}
\subfigure[]{
\includegraphics[width=0.4\textwidth]{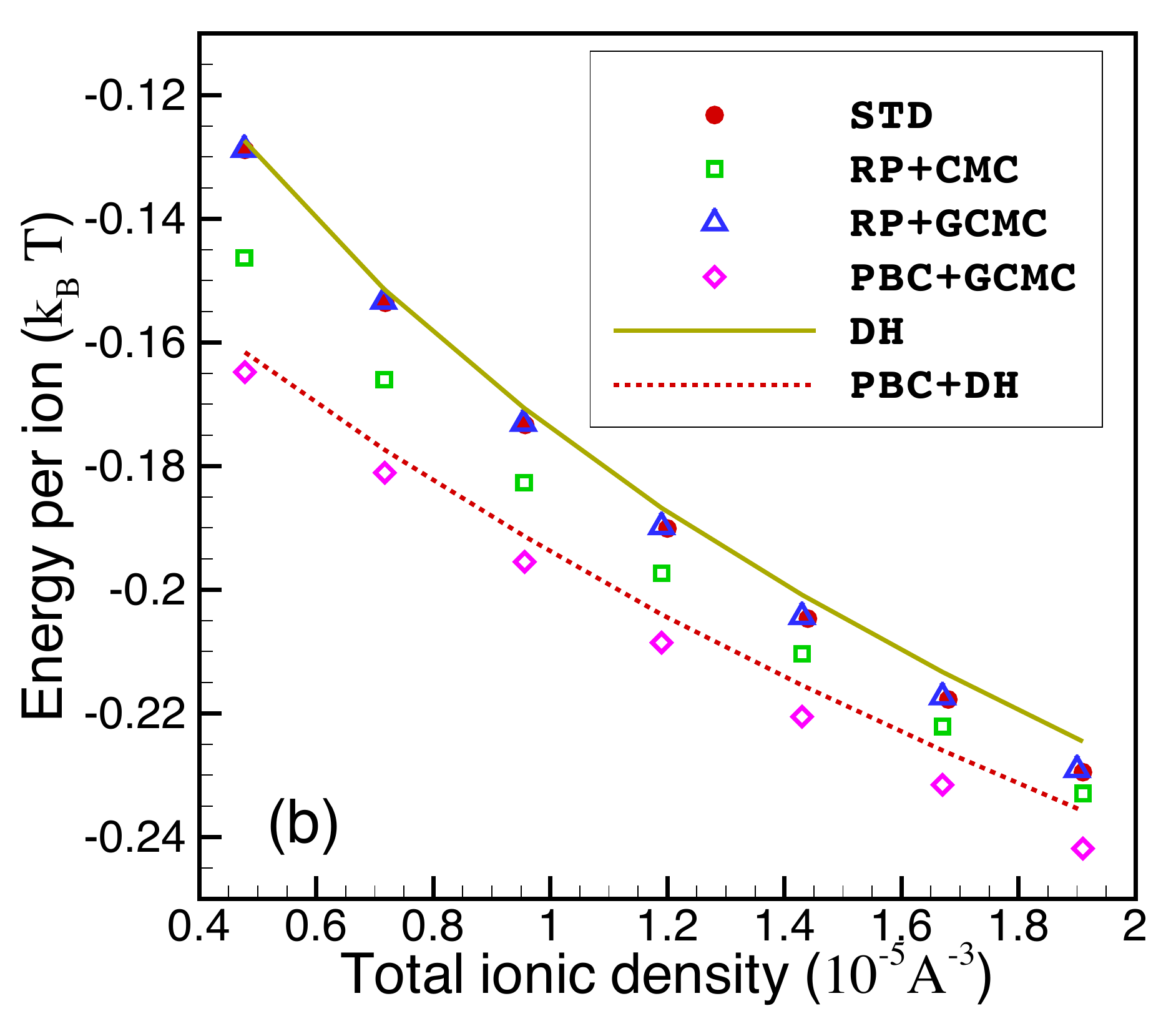}
\label{Fig_internal-energy-1}
}
\caption{(a) Screening charge density around a fixed ion, as a function of distance to the source.  Our multi-scale MC method (dark triangle, RP+GCMC) agrees the standard results (red dot, STD) to a high precision, indicating that it correctly captures the physics of an infinite size system.  By contrast, simulation using PBC (pink diamond, PBC+GCMC) gives substantially different results.  
The straight line is prediction by LPB in open space, Eq.~(\ref{screening-charge}), whereas the dashed curve is the prediction by LPB with PBC, Eq.~(\ref{rho_q-PBC}).  
System parameters in these simulations are listed in Sec.~\ref{sec:GCMC}.   (b) The average energy per ion, in the unit of $k_{\rm B}T$, as a function of ion density.  Again our multi-scale MC method agrees with the standard results to a high precision, whilst simulation using PBC yields largest errors among all methods.  The solid line (DH) is the prediction of {Debye-H\"uckel theory} in open space (the first term in RHS of Eq.~(\ref{chi-hardsphere})), and the dashed line, {Debye-H\"uckel theory} with PBC (Eq.~(\ref{chi-hardsphere})).  }
\label{Fig_comparison-1}
\end{figure*}

By periodically replicating the system, one introduces an infinite array of image charges for each charge belonging to the system under study.  These images are {\em unphysical} and break the {\em rotational symmetry}.   Furthermore, the total charge inside the simulation box must vanish,  for otherwise the summation over images would diverge.  This leads to an {\em unphysical} constraint of charge neutrality inside a {\em finite volume}.  In Fig.~\ref{Fig_screening-rho}, we show the average charge density around a fixed ion, obtained using different simulation methods.  Whilst large scale MC simulation (STD) gives a clean form of screened Coulomb, small scale simulation (with system size three Debye lengths) using PBC (PBC+GCMC) gives substantially different results.  The deviation, which grows with the distance to the test ion, is caused by the unphysical image charges.  Similarly, internal energies calculated using PBC also deviate substantially from the standard results, shown in Fig.~\ref{Fig_internal-energy-1}.  These artifacts are quantitatively less important for larger systems, and become invisible for infinite systems.   In principle, one can always simulate sufficiently large systems so that these artifacts become insignificant, or correct these artifacts for every microscopic configurations during the simulation.   This however generically leads to waste of computational resources, or slow-down of simulation processes. 

\begin{figure}[th!]
\includegraphics[width=0.48\textwidth]{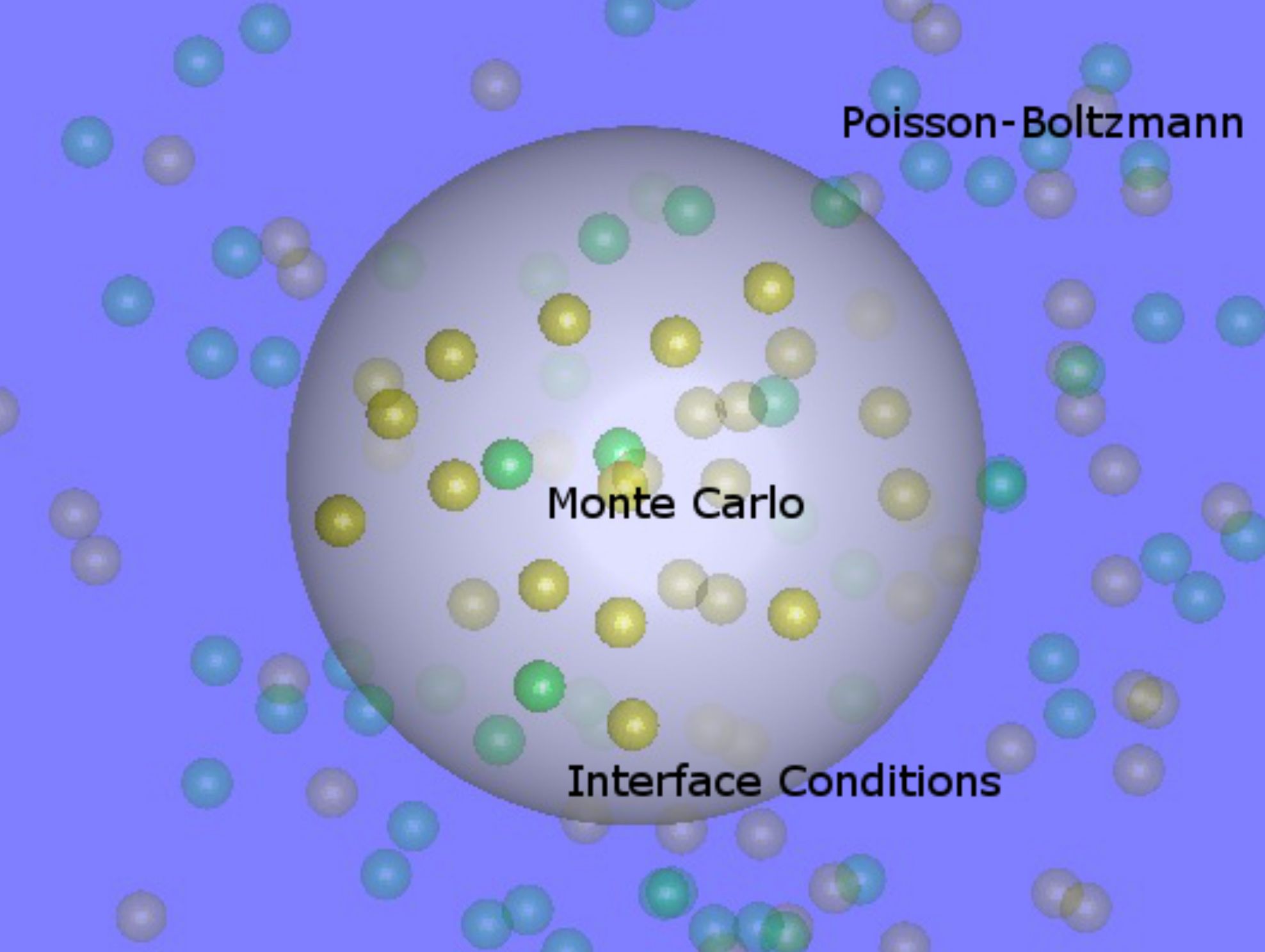}
\caption{Schematic illustration of the multi-scale reaction potential grand canonical Monte Carlo.  If this simulation method work well, the boundary of simulation cavity must behaves as a virtual one, i.e., the simulated system must behave as ions inside a virtual spherical domain in an infinitely large system. }
\label{Fig_schematics}
\vspace{-5mm}
\end{figure}

As an alternative to PBC, one may use a linear continuum theory to model the influence of the subsystem outside the simulation domain, whilst particles inside the domain are simulated explicitly.  For a dipolar system, the subsystem outside responds to the dipoles inside the cavity and exerts a field on the latter, which are conventionally called the {\em reaction field}.  The term {\em reaction field boundary condition} is named after Onsager \cite{Onsager:JACS:1936}, who first used this method to calculate the dielectric constant of dipolar fluids.  Similar multi-scale ideas, i.e.  treating dipolar molecules in near and far fields using different methods, however can be dated back to Clausius \cite{Clausius-dielectrics} and Mossotti \cite{Mossotti-dielectrics}.  For a discussion, see the monographs on dielectrics by Fr\"ohlich \cite{Frohlich-dielectrics}, and by Battcher \cite{Battcher-dielectrics}.  Born \cite{Born:ZP:1920} also used a similar idea to calculate the solvation (free) energy of ions in solvent.   We note that in the study of ionic systems, the term {\em reaction potential} would be more appropriate than {\em reaction field}, since it is the electrostatic potential, rather than field, that couples to the charges.   The same method can be generalized to molecules with arbitrary charge distributions.  

Simulation methods using multi-scale strategy have been studied by many authors \cite{Watts-1974,Barker1973mp,Barker:MP:1994,TSS+:JCP:1995,KW:JCP:1989,Alper-1993,BR:JCP:1994,WH:JPC:1995,Leeuw-1986-review}.  These methods were designed for study of either dipolar systems or ionic systems.  The simulation cavity can be either fixed and common for all simulated particles, or can be moving and individual for each particle.  The subsystem outside the cavity is always assumed to obey Poisson equation (for dipolar systems) or the linearized Poisson-Boltzmann equation (for ionic systems).  Finally, it appears that in most of the previous works, the reaction-field modeling is used in molecular dynamics simulations (MD), even though in principle, it is applicable for Monte Carlo simulations as well.  
% \xing{Add earlier works on Monte Carlo.}
 %\cite{Barker1973mp,Barker:MP:1994,TSS+:JCP:1995},  
 
To implement the multi-scale simulation, one must solve the continuum theory for each simulation step.  The classical Kirkwood expansion \cite{TK:JACS:1957}, though straightforward, is not efficient because of its slow convergence.  The image charge method, which was first discovered by Neumann \cite{Neumann:TL:1883} for polar systems, and have been generalized to ionic systems \cite{CDJ:JCP:2007,LBD+:JCP:2009,LBS+:JCP:2011}, transforms the Kirkwood expansion into a line integral, and therefore can be efficiently computed.  Recently, we improved the image method \cite{XLX:SJAM:2013} (using Mellin transform technique) so that it is applicable for Poisson-Boltzmann theory.
% and particularly for the case where the dielectric constant is identical inside and outside the cavity, which is the case of our multi-scale reaction-potential MC simulation of electrolytes.   

Regardless of the rich history, there are still a few contributions that we can make to improve this multi-scale simulation strategy.  Firstly, in order to mimic an infinite system using a finite and simulation domain (which is the essential task of almost all computer simulations), the total charge inside the simulation cavity must be allowed to fluctuate.  This means that the numbers of positive and negative ions must be allowed to fluctuate independently.  A ``grand canonical Ewald'' simulation where the ions are inserted and deleted in neutral pairs can not capture the physics of charge fluctuations.   In this work, we shall discuss in detail the artifacts due to suppression of charge fluctuations.   Secondly, all previous works assume without proof that the subsystem outside the cavity is described by the linearized Poisson-Boltzmann theory (LPB) for ionic systems.  We shall derive the correct linearized continuum theory and show that it reduces to the na\"ive LPB only in the dilute limit. For non-dilute electrolytes, the continuum theory may be considerably complicated.  Furthermore, for asymmetric electrolytes, there is also a constant {\em cavity potential} that acts on all ions inside the cavity.  Thirdly, we shall combine the multi-scale GCMC with a recently developed image charge method \cite{XLX:SJAM:2013} to simulate symmetric electrolytes.   We find that multi-scale simulation of a small system (with linear size of only three Debye length) is capable of capturing the physics of an infinite system.  For example, as shown in Fig.~\ref{Fig_screening-rho} and Fig.~\ref{Fig_internal-energy-1}, both the screening charge density around a fixed ion and the average internal energy per ion computed using our multi-scale GCMC (RP+ GCMC) agrees remarkably well with the standard results (STD).  By contrast, to achieve similar precision in simulations using PBC, the system would have to be at least as large as ten times Debye length. 

The remaining of this paper is organized as follows.  In Sec.~\ref{artifacts-PBC}, we analytically discuss LPB with PBC, and demonstrate two sources of artifacts of PBC, i.e., periodic image charges, and constraint of charge neutrality inside each unit cell.  In Sec.~\ref{sec:theory}, we start from the basic principle of statistical mechanics, and systematically derive the multi-scale MC simulation strategy.  This derivation made it clear that one should always use grand canonical ensemble for the subsystem inside the cavity, and that, for asymmetric electrolytes, there is a constant cavity potential acting on every ions inside. In Sec.~\ref{sec:implementation} we outline the numerical implementation of the multi-scale strategy using grand canonical MC simulation, as well as that of other simulation methods that are used for comparative studies.  In Sec.~\ref{sec:results}, we  simulate the symmetric primitive model, and compare the simulation results using different methods.  We demonstrate how our multi-scale method captures correctly the correlation effects between ions, whilst the Ewald summation methods fail to do so.  Finally in Sec.~\ref{sec:conclusion} we draw the conclusion remarks.

To simulate large systems, one need to use fast algorithms based on multipole expansion, such as the oct-tree algorithm \cite{Gan2014scm}, to expedite the computation of electrostatic energy.  A Monte Carlo simulation strategy that combine multi-scale modeling, grand canonical ensemble, image charge techniques, and oct-tree algorithm speeding-up can achieve a computational complexity of order of $N \log N$ for one Monte Carlo cycle, and is free of artifacts due to PBC.  It therefore constitutes a competitive alternative to molecular dynamics simulation methods for electrolyte systems.  This will be discussed in a future work.  

\section{Artifacts due to Periodic Boundary Conditions }
\label{artifacts-PBC}
%We discuss the artifacts due to periodic boundary conditions when it is applied to the study of electrolytes.  Generalization to asymmetric electrolytes is rather straightforward.  

Consider $N$ positive ions $q$ and $N$ negative ions $- q$ inside the simulation cube, which is centered at the origin and has length $L$, and volume $V = L^3$.   To simplify the analysis, we shall only focus on symmetric electrolytes in the dilute regime, where linearized Poisson-Boltzmann theory is applicable.  We shall for the moment assume that all ions that point-like, and consider the case of finite ion sizes towards the end of this section.  We periodically replicate the system, so that both the charge distribution and the electrostatic potential become periodic functions. 
%In accordance with the periodic boundary condition, the microscopic potential $\varphi(\rv)$ must also be periodic.  

We fix a positive ion $q$ at the origin, and let all other ions fluctuate according to the equilibrium Gibbs measure.  We shall calculate two quantities: 1) the screening charge density $\rho_q(\rv)$ around the test ion, and 2) the correlation potential $\psi$, i.e., the mean potential acting on the test ion due to all other ions.  These two quantities are calculated in the classical Debye and H\"uckel theory (DH) \cite{Debye-charging} for an infinite electrolyte.  What we are doing here is to generalize DH to a finite system with PBC.  The total internal energy of the system is related to $\psi$ via
\ba
E &=& \frac{1}{2} \sum_i q_i \sum_{j  \neq i}
\left\langle
\frac{q_j}{4 \pi \epsilon |\rv_i - \rv_j |} \right\rangle
%= \frac{1}{2} \left(2N \right) q \chi 
= N q \psi.
%\nonumber\\
\ea

Now according to the linearized Poisson-Boltzmann theory (LPB), the average number densities of {\em mobile} positive and negative ions are related to the mean potential $\phi(\rv)$ via the Boltzmann factor:
\begin{subequations}
\label{dist}
\ba
c_+(\rv) &=& \tilde{c}_+ e^{-\beta q \phi(\rv)}
\approx \tilde{c}_+ -\beta q \tilde{c}_+ \phi(\rv),
\\
c_-(\rv) &=& \tilde{c}_- e^{ \beta q \phi(\rv)} \approx \tilde{c}_-
 +\beta q \tilde{c}_- \phi(\rv),
\ea
\end{subequations}
where we have linearized the exponentials in the second steps.  For a physical system with infinite size, one usually requires that the mean potential $\phi(\rv)$ vanishes in the infinity.  Then according to Eqs.~(\ref{dist}), $\tilde{c}_{\pm}$ are the number densities of positive/negative ions in the infinity, i.e., the bulk ion densities \footnote{One might be surprised that the physical significance of parameters $\tilde{c}_{\pm}$ should depend on such a seemingly arbitrary requirement.  To see that this is natural, we note that the mean potential $\phi(\rv)$ and the mean charge density $\rho_q(\rv)$ are related by the Poisson Equation:
$- \Delta \phi(\rv) = \rho_q(\rv)/{\epsilon}$.  Note that the potential is not completely determined by the charge density.  That is to say, one must impose some condition in order to uniquely determine $\phi(\rv)$ for a given $\rho(\rv)$.  A choice of this condition is equivalent to an assignment of physical meanings to the parameters $\tilde{c}_{\pm}$ in Eqs.~(\ref{dist}).}.  

This argument is no longer applicable if periodic boundary condition is imposed, since the mean potential $\phi(\rv)$ does not vanish as infinity.  For periodic systems, the most convenient choice is that the mean potential $\phi(\rv)$ vanishes identically when integrated over the unit cell:
\be
\int_{\rm cell}\phi(\rv)d \rv=0.
\label{Gauge1}
\ee
This is equivalent to the ``conductor boundary condition'' that are popularly used in Ewald summation method, which requires that the Fourier transform of electrostatic potential does not have a $\mathbf{k} = 0$ component.  Now if we integrate Eqs.~(\ref{dist}) over the unit cell with volume $V$, the LHS are the total numbers of mobile positive/negative ions in the unit cell, which are $N -1$, and $N$, respectively, whereas the RHS are  $\tilde{c}_{\pm} V$ respectively.  This allows us to determine the parameters $\tilde{c}_{\pm}$ as
\ba
\tilde{c}_+ = \frac{N - 1}{V}, \quad
\tilde{c}_- =  \frac{N}{V}.
\label{tilde_c_pm}
\ea

% In fact, we should identify $\tilde{c}_{\pm}$ to be the number density of positive/positive ions in the special case where the mean potential is zero.
% We need to realize that the total charges inside a unit cell must vanish, otherwise, the electrostatic energy would diverge.  Hence the total charges of all mobile ions $(N_+ ) $
%:

Let us now invoke the Poisson equation, use Eqs.~(\ref{dist}), and expand to the first order in $\phi$.  We obtain
\ba
- \Delta \phi(\rv) &=& \frac{1}{\epsilon} \rho_q(\rv)
%\nonumber\\
=   \frac{1}{\epsilon} q \, \sum_{\lv} \delta(\rv - \lv)
+ q \left( c_+(\rv) -  c_-(\rv) \right)
\nonumber\\
&=&  \frac{1}{\epsilon} q \, \sum_{\lv} \delta(\rv - \lv)
- \frac{q}{\epsilon \, V} - \tilde{\kappa}^2\phi,
\label{GPB}
\ea
where $\lv=(L_x n_x, L_y n_y, L_z n_z)$ are lattice vectors, $n_x$, $n_y$, $n_z$ are integers, and $\tilde{\kappa}$ is the inverse Debye length: 
\ba
\tilde{\kappa}
%&=& \sqrt{\beta q^2 (\tilde{c}_+ + \tilde{c}_- )/\epsilon}
= \sqrt{\epsilon^{-1} \beta \, q^2 (2 N - 1)/V }.
\label{kappa-PBC}
\ea
The first term in RHS of Eq.~(\ref{GPB}) is due to the test ion and its periodic images, whilst the second term corresponds to a {\em uniform and negative} charge density, arising due to the fact {\em there is one more negative mobile ion than positive mobile ions}, see Eq.~(\ref{tilde_c_pm}).  {\em This in turns originates from the artificial constraint of charge neutrality inside each primitive cell. } Integrating both sides over a unit cell, one easily see Eq.~(\ref{Gauge1}) is indeed satisfied.  
Note that inverse Debye length $\tilde{\kappa}$ in Eq.~(\ref{kappa-PBC}) is also slightly different from its usual form in free space. 

Eq.~(\ref{GPB}) can be easily solved subject to PBC:
\be
\phi(\rv)=
              \sum\limits_{\lv}
              \frac{q \, e^{-\tilde{\kappa} |\rv-\lv|}}
              {4\pi\epsilon|\rv-\lv|}
              - \frac{q}{\epsilon \tilde{\kappa}^2V},
\label{Res1}
\ee
which can be easily shown to satisfy the condition Eq.~(\ref{Gauge1}).  Taking the Laplacian of this equation, we find the total average charge density:
\be
\rho_q(\rv) = q \, \sum_{\lv} \delta(\rv - \lv)
- \sum_{\lv}\frac{q \tilde{\kappa}^2
\,  e^{-\tilde{\kappa} |\rv-\lv|}}{4\pi |\rv-\lv|},
\label{rho_q-PBC}
\ee
where the first term is due to the test ion and its images, whilst the second term is sum of the screening charge densities around these ions.  Eq.~(\ref{rho_q-PBC}) (averaged over all orientations of $\rv$) is plotted in Fig.~\ref{Fig_screening-rho} as PBC+DH, which agrees remarkably well with MC simulations with PBC (PBC+GCMC).  

The correlation potential $\psi$ acting on the fixed ion can be easily obtained by subtracting off the part due to the test ion itself from Eq.~(\ref{Res1}), and take the local limit $\rv \rightarrow 0$:
\ba
\psi
&=& \lim_{\rv \to 0}\left[ \phi(\rv)-\frac{q}{4\pi \epsilon |\rv| }\right]
\nonumber\\
&=& - \frac{q \, \tilde{\kappa}}{4\pi\epsilon}
 +\sum\limits_{\lv\neq\vec{0}}
 \frac{q \,  e^{-\tilde{\kappa} |\lv|}}{4\pi\epsilon|\lv|}
              - \frac{q }{\epsilon \tilde{\kappa}^2 V}.
              \label{chi-point}
\ea
The first term is just the correlation potential in free space as predicted by Debye-H\"uckel theory, and the other two terms constitute the artifacts due to periodic boundary conditions, both of which vanish as the system size $L$ goes to infinity (with the Debye length fixed). 

If the ions are not point-like, but are hard spheres with diameter $d$ (the primitive model), the above results need to be modified properly.  The resulting theory is considerably more complicated.  If the ion density is not very high, however, we may consider corrections due to hardcores as small perturbations, same as the artifacts due to PBC.  Then all we have to do is to correct the first term in Eq.~(\ref{chi-point}), i.e., replace it by the correlation potential of a hard sphere ion, which was worked out by Levin and Fisher some time ago \cite{Levin-charge-review, Fisher-Coulomb-review}.  This leads to
\be
\psi =
-\frac{q \, \tilde{\kappa}}{4\pi\epsilon(1+\tilde{\kappa} a)}
 + \sum\limits_{\lv\neq\vec{0}}
 \frac{q \, e^{-\tilde{\kappa} |\lv|}}{4\pi\epsilon|\lv|}
               - \frac{q}{\epsilon \tilde{\kappa}^2V}.
                             \label{chi-hardsphere}
\ee
This result is plotted in Fig.~\ref{Fig_internal-energy-1} (PBC+DH), which agrees reasonably well with the MC simulations with PBC if the ion density is low.  The agreement becomes increasingly worse as the density increases, indicating that the approximation underlying Eq.~(\ref{chi-hardsphere}) becomes increasingly inaccurate.   

In Fig.~\ref{Fig_comparison-1}, we have chosen the size of simulation box to be approximately three Debye length, in order to highlight the artifacts due to PBC.  In many simulations, the size of simulation box is chosen to be much larger than the Debye length.  In this case, the second term in Eq.~(\ref{chi-hardsphere}) is exponentially small comparing with the third term.  The artifacts is therefore completely due to the last term.  The relative error (in the dilute regime $\tilde{\kappa} a \rightarrow 0$) due to PBC is the given by 
\be
\frac{\delta \psi}{\psi} \approx \frac{4 \pi}{(\tilde{\kappa} L )^3}. 
\ee
To make this relative error less than 1\%, one would have to choose the size of simulation box to be larger than ten times Debye length.  This may be very difficult to achieve in many simulations.  By strong contrast, using the multi-scale GCMC simulation method, we can faithfully simulate an infinite system using a simulation cavity with radius only three times Debye length.

\section{Multi-scale Modeling}
\label{sec:theory}
%\begin{comment}

%\end{comment}

\subsection{Effective Interaction}
We shall study the {\em primitive model} of electrolytes, where the solvent (water) is modeled implicitly as a dielectric medium with relative permittivity $\epsilon/\epsilon_0 = 80 $, whilst ions are modeled as hard spheres with a point charge at the center.  Furthermore, the permittivity of the hard spheres is assumed to be the same as that of the solvent.  We introduce a spherical {\em simulation cavity}, which is already schematically illustrated in Fig.~\ref{Fig_schematics}.  The subsystem $\mathcal{A}$ {\em inside the cavity} consists of $N$ charges $\{q_1,\ldots,q_N\}$ with positions denoted by ${\mathbf X}=\{\xv_1,\ldots, \xv_N\}$, whereas the subsystem $\mathcal{B}$ {\em outside the cavity} consists of $M$ charges $\{\theta_1,\ldots, \theta_M\}$ with positions denoted $\Yv = \{\yv_1,\ldots, \yv_M\}$ respectively.
%:
%We shall only study symmetric electrolytes in this work, hence $q_i, \theta_i = \pm q$.
%:

The total Hamiltonian of the overall system consists of three parts:
\be
H = H^{\mathcal A} + H^{\mathcal B} + H^{\mathcal{AB}},
\label{H-B-AB}
\ee
where $H^{\mathcal A}$ and $H^{\mathcal B}$ are the Hamiltonian of isolated subsystems ${\mathcal A}$ and ${\mathcal B}$:
\begin{subequations}
\ba
H^{\mathcal A} &=& \sum \limits_{i < j }^{N}
\left[
\frac{1}{4\pi\epsilon}
\frac{q_i q_j} {|\mathbf{x}_i-\mathbf{x}_j |}
+w(\xv_i - \xv_j)
\right],
\label{H_A}\\
H^{\mathcal B} &=&
 \sum \limits_{ \alpha <  \beta }^{M}
\left[
\frac{1}{4\pi\epsilon}
\frac{\theta_{\alpha} \theta_{\beta}}
{|\yv_{\alpha}-  \yv_{\beta} |}
+w( \yv_{\alpha}-  \yv_{\beta}  )
\right] .
\ea
\end{subequations}
The first term in each bracket is the long-range Coulomb interaction and the second term $w(\xv - \yv)$ is the short range interaction, which is assumed to be independent of the ion species .   For the primitive model, $w(\xv - \yv)$ is just the pairwise hardcore repulsion:
\be
w(\xv - \yv) = \left\{ \begin{array}{ll}
0, & |\xv - \yv| > d;
\vspace{3mm}\\
\infty, & |\xv - \yv| < d.
\end{array}\right.
\ee
The last part in the RHS of Eq.~(\ref{H-B-AB}), $H^{\mathcal{AB}}$, represents the interaction between subsystems $\mathcal{A}$ and $\mathcal{B}$:
\begin{equation}
\label{H_AB}
H^{\mathcal{AB}} = 
\sum\limits_{i=1}^{N}\sum\limits_{\alpha=1}^{M}
\left[
\frac{1}{4\pi\epsilon}
\frac{q_i\theta_\alpha}{|\mathbf{x}_i-\mathbf{y}_\alpha|}
+ w(\mathbf{x}_i-\mathbf{y}_\alpha)
\right]. 
\end{equation}
We note that the short range part of $H^{\mathcal{AB}}$ in Eq.~(\ref{H_AB}) becomes nonzero only if the distance between $\xv_i$ and $\yv_{\alpha}$ is smaller than $d$, which can happen only if both particles are very close to the cavity boundary, hence it has no influence on the ions that are at least one ion-diameter away from the boundary.  We shall therefore ignore this short range interaction between $\mathcal{A}$ and $\mathcal{B}$.  Accordingly, we shall exclude a thin spherical shell near the cavity boundary for collection of simulation data.   $H^{\mathcal{AB}}$ can then be written into the following form:
\be
\label{H_AB-1}
H^{\mathcal{AB}} = \sum_{i = 1}^N q_i \, \varphi(\xv_i, \Yv)
\ee
where $\varphi(\xv_i, \Yv)$ is the {\em potential acting on the charge $ q_i$ at $\xv_i$ due to all ions in $\mathcal{B}$}, in the specific micro-configuration $\Yv = \{\yv_1, \ldots, \yv_M\}$:
\be
\varphi(\xv_i, \Yv) = \sum\limits_{\alpha=1}^M
\frac{1}{4\pi\epsilon}
 \frac{\theta_{\alpha}}{|\xv_i - \yv_\alpha|}. 
 \ee

The {\em grand canonical partition function} of the overall system can be written as
\ba
\Xi = \Tr_{\mathcal A} \Tr_{\mathcal B} \,
e^{-\beta(H^{\mathcal A} + H^{\mathcal B} + H^{\mathcal{AB}})}, 
\label{Partition-total}
\ea
where the traces $\Tr_{\mathcal A}$ and $\Tr_{\mathcal B}$ represent integrations of all the variables as well as summation over particle numbers:
\ba
\Tr_{\mathcal A} &=& \sum\limits_{\{N_s\}}
\left( \prod_{s} \frac{ e^{ N_s \beta \mu_s}}
{N_s! \Lambda_s^{3N_s} } \right)
        \int_  A d^{3N}\mathbf{X},   \\
\Tr_{\mathcal B} &=& \sum\limits_{\{M_s\}}
\left( \prod_{s} \frac{ e^{ M_s \beta \mu_s}}
{M_s! \Lambda_s^{3M_s} } \right)
        \int_B d^{3M}\mathbf{Y}.
\ea
Here $N_s$ and $M_s$ are the numbers of mobile particles of $s$-th species  in subsystems A an B respectively, whereas $\mu_{s}$ are their chemical potentials, and $\Lambda_s$ are some microscopic length scales. 

We integrate out the variables in subsystem ${\mathcal B}$, and obtain an effective theory for the subsystem ${\mathcal A}$:
\begin{subequations}
 \label{Z-effective-A}
\ba
\Xi &=& \Tr_{\mathcal A} \, e^{-\beta H^{\mathcal A}}
\left(
\Tr_{\mathcal B} \, e^{-\beta\left(H^{\mathcal B}
+ H^{\mathcal{AB}}\right)}
\right)  \nonumber\\
 &\equiv& \Tr_{\mathcal A} \, e^{-\beta
 \left(H^{\mathcal A} +\delta H^{\mathcal A} \right)}
 \equiv  \Tr_{\mathcal A} \, e^{-\beta H^{\mathcal A}_{\rm eff}},
 \label{Z-effective-A-1} 
\ea
where $\delta H^{\mathcal A}$ represents the additional effective interactions of subsystem ${\mathcal A}$ as mediated by subsystem ${\mathcal B}$:
\begin{equation}
e^{-\beta \delta H^{\mathcal A} } =  \Tr_{\mathcal B}
\, e^{-\beta \left(H^{\mathcal{AB}} + H^{\mathcal B}\right)}.
\label{rf_def}
\end{equation}
\end{subequations}
Therefore we can study the subsystem ${\mathcal A}$ {\em only} with an effective Hamiltonian
\ba
H^{\mathcal A}_{\rm eff} = H^{\mathcal A} + \delta H^{\mathcal A}
= H^{\mathcal A} - \beta^{-1} \Tr_{\mathcal B} \,
e^{-\beta \left(H^{\mathcal{AB}} + H^{\mathcal B}\right)},
\label{HA_eff-def}
\ea
and the influences of the subsystem ${\mathcal B}$ on ${\mathcal A}$ are automatically taken care of.  We will find that $\delta \! H^{\mathcal A}$ can be approximately calculated and the result assumes a simple and physically transparent form.  

We emphasize that Eqs.~(\ref{Z-effective-A-1}) and (\ref{Partition-total}) are {\em grand canonical partition functions}.  Hence the ion numbers for each species in ${\mathcal A}$ and ${\mathcal B}$ are all stochastic variables.  Now it is well known that for large systems (such as subsystem ${\mathcal B}$), all ensembles are equivalent to each other.  But for small systems (such as ${\mathcal A}$), different ensembles are inequivalent.   It is therefore important to use grand canonical ensemble, rather than canonical ensemble for ${\mathcal A}$ in our multi-scale modeling.  We shall demonstrate this point by comparing simulations results using GCMC and CMC in Sec.~\ref{sec:results}.  

%The Coulomb many body systems that we can simulate are certainly of very small size.

%Following the terminology of {\em reaction field} by Onsager \cite{Onsager-RF}, we shall call $\delta H^{\mathcal A} $ {\em the reaction potential energy}.

\subsection{Method of Debye Charging }
To calculate the effective interaction $\delta H^{\mathcal A}$, we shall use the Debye {\em charging method} \cite{Debye-charging}.  Let us scale charges of all ions $q_i$ in subsystem $\mathcal{A}$ by a factor $\lambda$:
\be
q_i \rightarrow \lambda \, q_i,
\ee
so that Eq.~(\ref{H_AB-1}) becomes
\ba
H^{\mathcal{AB}}_{\lambda}  = 
\sum_{i = 1}^N \lambda \, q_i \, \varphi(\xv_i, \Yv)
= \lambda \, H^{\mathcal{AB}}
\label{intersystem}
\ea
  Eq.~\eqref{rf_def} then becomes:
\begin{equation}
e^{-\beta \delta H_{\lambda}^{\mathcal A} } =  \Tr_{\mathcal B}
\, e^{-\beta \left(  \lambda \, H^{\mathcal{AB}} + H^{\mathcal B}\right)}.
\label{rf_def-lambda}
\end{equation}
Taking  derivative of Eq.~(\ref{rf_def-lambda}) with respect to $\lambda$, we find
\begin{equation}
\frac{d}{d\lambda} \delta H^{\mathcal{A}}_{\lambda}
= \sum\limits_{i=1}^{N}
q_i \big\langle \varphi(\xv_i, \Yv) \big\rangle_\lambda,
\label{d_rf}
\end{equation}
 where the $\lambda$-dependent average $\langle O \rangle_\lambda$ is defined as
\begin{equation}
\langle O \rangle_\lambda = \frac{\Tr_{\mathcal B} \, O \,
e^{-\beta( H^{\mathcal{B}} +  \lambda \, H^{\mathcal{AB}} )}}
{\Tr_{\mathcal B} \, e^{-\beta( H^{\mathcal{B}} +  \lambda \, H^{\mathcal{AB}} )}}.
\label{exp_def1}
\end{equation}
The physical significance of $\big\langle \varphi(\xv_i, \Yv) \big\rangle_\lambda$ is therefore {\em the average potential at $\xv_i$ (inside the cavity, occupied by an ion $\lambda q_i$) due to all ions in ${\mathcal B}$. }  Note that the average is over the statistical fluctuations of all ions outside, with the locations of all ions $\lambda q_i$ inside fixed.  To calculate this average, we treat $\lambda$ as a small parameter and use the {\em linear response theory}.  That is, we expand Eq.~\eqref{exp_def1} to the first order of $\lambda$:
\ba
\langle O \rangle_\lambda
&\approx&
\langle O  \rangle_0
- \lambda \beta
\left(
\left \langle O
H^{\mathcal{AB}}
\right\rangle_0
- \left \langle O \right\rangle_0
 \left \langle H^{\mathcal{AB}}
\right\rangle_0
\right).
\nonumber\\
&+& O(\lambda^2)
\label{cumu_exp}
\ea
Note that the average $\langle O \rangle_0$ in RHS is defined by Eq.~(\ref{exp_def1}) with $\lambda$ set to zero, i.e., with the interaction between two subsystems switched off:
\begin{equation}
\label{O-ave-def}
\langle O \rangle_0
= \frac{\Tr_{\mathcal B} \,O \,
e^{-\beta \left( H^{\mathcal{B}} 
+ H^{\mathcal{AB}}\right)
} }
{\Tr_{\mathcal B} \,
e^{-\beta \left( H^{\mathcal{B}} 
+ H^{\mathcal{AB}} \right)
}}.
\end{equation}

Let us define a kernel $\chi(\rv,\rv')$ in terms of the connected correlation function of $ \varphi(\rv, \Yv)$ as follows:
\ba
\label{chi-def}
  \chi(\rv,\rv') =   &-& \beta  \big[  
  \left \langle \varphi(\rv, \Yv)   \varphi(\rv', \Yv)  \right \rangle_0
  \nonumber\\
&  - &  \left \langle  \varphi(\rv, \Yv)  \right \rangle_0
  \left \langle  \varphi(\rv', \Yv) \right\rangle_0
  \big] . \nonumber
\ea
Setting $O = \varphi(\xv_i, \Yv)$ in Eq.~(\ref{cumu_exp}) and using Eqs.~(\ref{O-ave-def}),  (\ref{H_AB-1}), we obtain:
\ba
\langle \varphi(\xv_i, \Yv) \rangle_\lambda
=  \langle \varphi(\xv_i, \Yv)  \rangle_0 +
\lambda \sum_j q_j  \chi(\xv_i,\xv_j).
\label{potential-decomposition}
\ea
The first term in the RHS is the average potential at $\xv_i$  {\em in the absence of any charges in } $\mathcal{A}$, generated by an isolated subsystem ${\mathcal B}$.   We shall call this term the {\em cavity potential} $\Phi_{\rm cav}(\rv)$, following the terminology of Onsager~\footnote{Onsager \cite{Onsager:JACS:1936} coined the term {\em cavity field}, because he was studying a dipole inside a cavity. }.  Evidently it satisfies the Laplace equation inside the cavity, and is invariant under arbitrary rotation.  It then follows that $\Phi_{\rm cav}(\rv)$ must be a constant in the whole cavity:
\be
\Phi_{\rm cav} = \left\langle \varphi(\rv, \Yv) \right \rangle_0.
\label{cavity-potential-def}
\ee
Furthermore, for symmetric electrolytes, the cavity potential vanishes identically due to charge inversion symmetry.  For asymmetric electrolytes, $\Phi_{\rm cav}$ is generally non-vanishing. 
This term has been missed by all previous works on reaction-field modeling of charged systems. 

The second term in the RHS of Eq.~(\ref{potential-decomposition}) is linear in $\lambda$.  It therefore arises due to the linear reaction of subsystem ${\mathcal B}$ to sources charges $\{\lambda \,q_1, \ldots, \lambda \,q_N\}$ in $\mathcal{A}$.  We shall therefore call it the {\em reaction potential}~\footnote{Onsager \cite{Onsager:JACS:1936} used the term {\em feaction field} to denote the electrostatic field acting on a dipole inside the cavity due to the reaction of the subsystem outside. }.  

Finally, combining Eqs.~(\ref{cavity-potential-def}) and (\ref{potential-decomposition}), and substituting them back into Eq. \eqref{d_rf},  integrating the latter over the charging parameter $\lambda$ from $0$ to $1$, we obtain the effective interaction of ${\mathcal A}$ mediated by ${\mathcal B}$: 
%\xing{We ignore $\delta H^{\mathcal A}_{\lambda = 0} $, which is due to depletion effects. }:
\ba
\delta H^{\mathcal A} =
\sum \limits_i q_i
\Phi_{\rm cav}
+ \frac{1}{2}\sum\limits_{i,j=1}^{N}
q_i q_j \chi ( \mathbf{x}_i, \mathbf{x}_j ). 
\label{rf_energy}
\ea
%Our task is therefore reduced to the calculation of the cavity potential $\Phi_{\rm cav}$ and the correlation function $\chi ( \mathbf{x}_i, \mathbf{x}_j )$.
	
The correlation function $\chi(\rv,\rv')$ is intimately related to the electrostatic Green's function of the subsystem ${\mathcal B}$.  To see this, let us insert a test ion with charge $q$ at $\rv'$ inside the otherwise empty cavity, and calculate the average total potential at $\rv$, which may be inside or outside the cavity.  This potential is the linear superposition of a part due to the source charge $q$, and another part due to all ions in  subsystem ${\mathcal B}$.  The later has the form given by Eq.~(\ref{potential-decomposition}), to the first order in $q$ (with $\lambda$ set to unity, of course).  Hence the total mean potential at $\rv$ is given by
\be
\Phi_{\rm cav}  + q\Bigg[  \frac{1}{4 \pi \epsilon |\rv - \rv'|}  + \chi(\rv,\rv') \Bigg]
+ O(q^2).
\ee
The sum inside the bracket describes the linear response of mean potential 
at $\rv$ due to a {\em unit point charge} at $\rv'$, and therefore is the {\em electrostatic  Green's function}:
\ba
G(\rv,\rv')
 =   \frac{1}{4 \pi \epsilon |\rv - \rv'|}
+ \chi(\rv,\rv').
\label{G-chi-relation}
\ea
Substituting this back into Eqs.~(\ref{HA_eff-def}) and (\ref{rf_energy}), we find the effective Hamiltonian $H^{\mathcal A}_{\rm eff}$:
\ba
H^{\mathcal A}_{\rm eff} 
&= & \sum\limits_{i < j}^{N}
q_i q_j G( \mathbf{x}_i, \mathbf{x}_j ),
\nonumber\\
&+&
\sum \limits_i  \left[ q_i \Phi_{\rm cav}
+  \frac{1 }{2}q_i^2 \, \chi ( \mathbf{x}_i, \mathbf{x}_i ) \right]
\label{H_A-eff}
\ea
where the first term is the effective interaction between different ions, and the second term is the self-energy of ions mediated by subsystem ${\mathcal B}$.  

\subsection{Linearized Poisson-Boltmann Theory}
The Green's function $G(\rv, \rv')$ encodes the linear response properties of the subsystem ${\mathcal B}$, that is an electrolyte with {\em a spherical void} (the empty cavity).  More precisely, it is the incremental mean potential at the field point $\rv$ due to the insertion of a unit test charge at the source point $\rv'$. If $\rv$ is inside the cavity, $G(\rv, \rv')$ obviously satisfies the Poisson equation:
\begin{subequations}
\be
- \Delta_{\rv} G(\rv, \rv') =  \frac{1}{\epsilon}
	\delta(\rv - \rv'), \quad \quad\quad \,\,\,
 	 \rv\in  \mathcal{A}.
	 \label{G-eq-A}
\ee
If $\rv$ is outside the cavity, it satisfies a linear integro-differential equation:
\be
- \Delta_{\rv} G(\rv, \rv') + \int \alpha(\rv, \rv'')  G(\rv'', \rv')d {\rv''} = 0,
 \quad \quad \rv \in  \mathcal{B}.
	 \label{G-eq-B-0}
\ee
The kernel $\alpha(\rv, \rv'')$ is generally unknown.  At the level of first order perturbation, however, we may approximate  $\alpha(\rv, \rv'')$ by the kernel that corresponds to a uniform electrolyte, which only depends on the difference $\rv - \rv''$, because of the translational symmetry.  The Fourier transform of this kernel was calculated analytically recently, both for symmetric and asymmetric electrolytes \cite{Dingmn-2015}.  The results are still quite complicated.  

In this work, we shall only study symmetric electrolytes in the dilute regime, where LPB is applicable in the bulk.  This entails two essential simplifications for the effective Hamiltonian Eq.~(\ref{H_A-eff}).  Firstly the cavity potential vanishes identically.  Secondly, the  kernel $\alpha(\rv, \rv')$ has the simple form $\kappa^2\, \delta (\rv - \rv')$, where $\kappa$ is the inverse (bare) Debye length, given by
\be
\kappa^2 = \epsilon^{-1} \beta \sum c_{\alpha} q_{\alpha}^2
= 2 \epsilon^{-1} \beta  c q^2,
\label{kappa-PB}
\ee
with $c$ the average ion density. Consequently, Eq.~(\ref{G-eq-B-0}) reduces to the linearized Poisson-Boltzmann equation:
\ba
- \Delta_{\rv} G(\rv,\rv') +  \kappa^2 \, G(\rv,\rv') = 0,
\quad \quad \rv\in \mathcal{B}.
	 \label{G-eq-B}
\ea
\end{subequations}
We shall deal with the cases of dense electrolytes and asymmetric electrolytes in a future work. 

We also need to determine the boundary conditions satisfied by the Green's function.  At infinity, it clearly satisfies the free boundary condition:
%:
\begin{subequations}
\label{G-BCs}
\begin{equation}
\lim \limits_{\rv \rightarrow \infty } G(\rv,\rv' ) \rightarrow  0.
\label{G-BCs-infty}
\end{equation}
On the cavity interface, $r = R$, $G(\rv,\rv')$ and its normal derivative are continuous:
\ba
\lim_{{r \rightarrow R^-}} G(\rv,\rv') &=&
\lim_{{r \rightarrow R^+}} G(\rv,\rv'),
\vspace{3mm} \\
\nonumber\\
\lim_{{r \rightarrow R^-}}
 \frac{\partial G(\rv, \rv')}{\partial r}
&=& \lim_{{r \rightarrow R^+}} 
 \frac{\partial G(\rv, \rv')}{\partial r},
\ea
\end{subequations}
where $r \rightarrow R^{\pm}$ mean that the field point $\rv$ approaches the interface from outside/inside, respectively. 

%\Yihao{Need to consider $\epsilon_R$ here. If the source charge is at origin, how large is $\chi$?}
%Finally, the reaction potential $\chi(\rv,\rv')$ is related to the Green's function via Eq.~(\ref{G-chi-relation}).

Our method therefore works as follows.  Firstly we find the Green's function by solving Eqs.~(\ref{G-eq-A}), (\ref{G-eq-B}), subject to boundary conditions (\ref{G-BCs}), then use Eq.~(\ref{G-chi-relation}) to find the reaction potential $\chi(\rv,\rv')$, and finally use Eq.~(\ref{H_A-eff}) to find the effective Hamiltonian.

\section{Numerical Implementation}
\label{sec:implementation}
In this section, we discuss the numerical implementation of our multi-scale MC method, as well as other simulation methods that we use for comparison.  

\subsection{Computation of the Green's Function}
In one of our previous works \cite{XLX:SJAM:2013}, we discussed an efficient image charge method for solving Eqs.~(\ref{G-eq-A}) and (\ref{G-eq-B}), subjected to boundary conditions Eqs.(\ref{G-BCs}).  Here we briefly summarize the results in order to make this work self-contained.  Let $\rv'$ and $\rv$ be, respectively, the source point and the field point.  We obtain the reaction potential $\chi(\rv, \rv')$ in the form of Kirkwood series \cite{TK:JACS:1957}: 
%(which is related to the Green's function via Eq.~(\ref{G-chi-relation})):
\be
\chi(\mathbf{r},\mathbf{r}') = \frac{1}{4\pi\epsilon R}
\sum\limits_{n=0}^{\infty}
\left( \frac{r r'}{R^2} \right)^n
M_n(u)P_n(\cos\theta),
\label{Green_rf_series}
\ee
where $\theta$ is the angle between $\mathbf{r}$ and  $\mathbf{r}'$, $P_n(x)$ the Legendre polynomials, $R$ the cavity radius, $u=\kappa R$, and
\be
M_n(u) = \frac{(n+1)k_n(u)+uk_n'(u)}{nk_n(u)-uk_n'(u)},
\ee
with $k_n(u)$ the modified spherical Hankel functions:
\be
k_n(u) = \frac{\pi e^{-u}}{2u}\sum\limits_{l=0}^n
			\frac{(n+l)!}{l!(n-l)!}\frac{1}{(2u)^l}.
\ee

%\begin{comment}
\begin{figure}[thb!]
\includegraphics[width=0.44\textwidth]{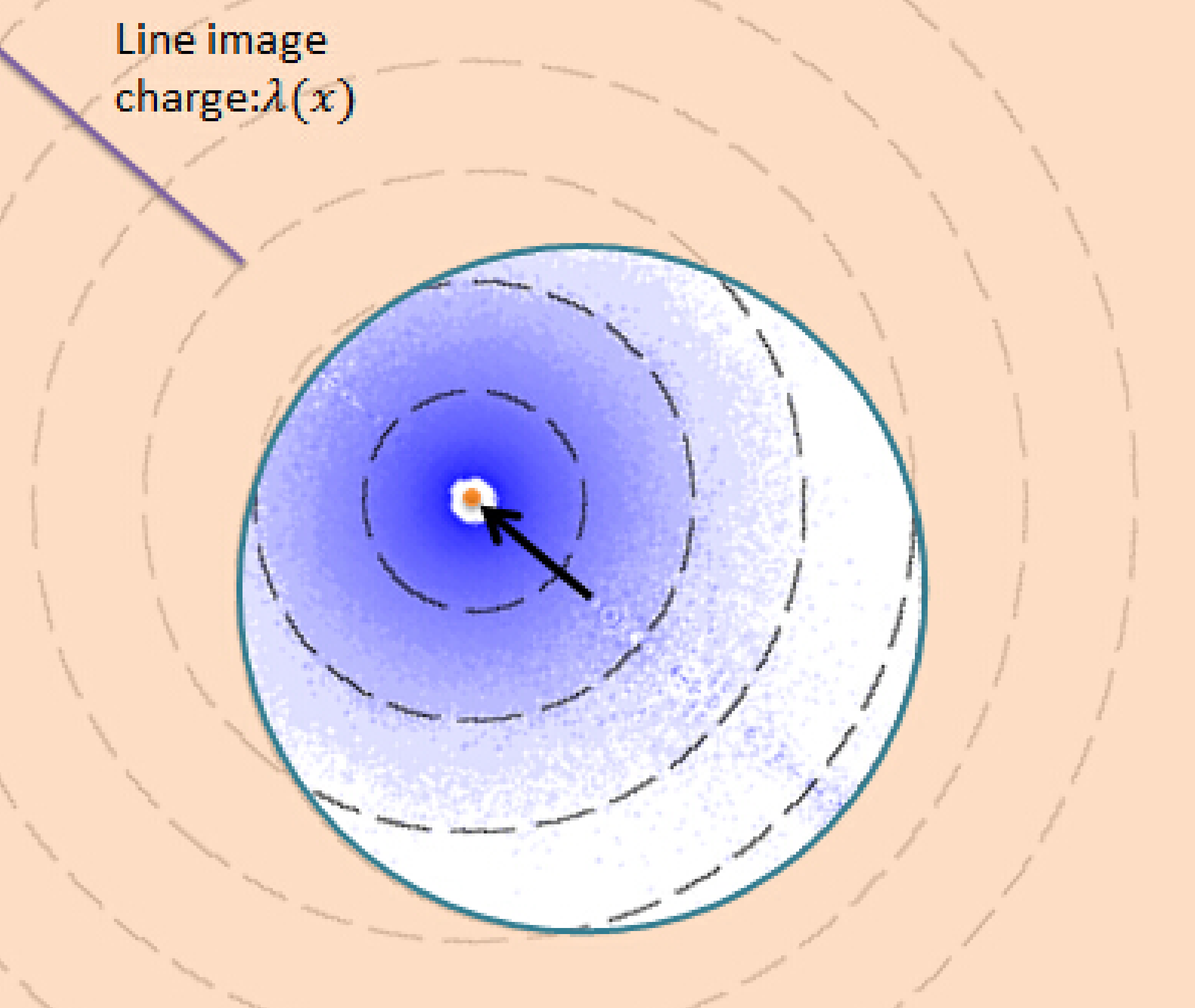}
\caption{Schematics. A source charge inside the cavity (red point), and the reaction potential  $\chi(\rv,\rv')$ generated by a line of image charges (blue straight line) as given by Eq.~(\ref{Green_rf_int_image}). The blue color inside the cavity is the density plot of screening charge density, obtained directly from simulation data.  The dashed circles are the contour lines as predicted by LPB.  The fact that these two plots agree with each other demonstrates the physics of an infinite system is faithfully reproduced in our multi-scale simulation using a small cavity.  }
\label{Fig_illustration}
\vspace{-4mm}
\end{figure}
%\end{comment}

Eq.~(\ref{Green_rf_series}) has a useful integral representation:
\be
\chi(\mathbf{r},\mathbf{r}') =
\frac{1}{4\pi\epsilon}\int_{r_{\! K}}^{\infty}
\frac{\lambda(\textit{t})}{|\mathbf{r}-\mathbf{t}|} dt,
\label{Green_rf_int_image}
\ee
where the vector ${\mathbf t}$  is parallel to $\rv'$ and can be written as
$${\mathbf t} = t  \left( \frac{{\mathbf r}'}{r'}\right)
= t \, \hat{{\mathbf r}}'. $$
Its magnitude $t$ runs from $r_{\! K} = R^2/r'$ to infinity.
Eq.~(\ref{Green_rf_int_image}) has an intuitive physical interpretation: it is the potential at $\rv$ due to a {\em fictitious} straight-line of image charges which starts at the Kelvin point  $\mathbf{r}_{\! K} =  {R^2\mathbf{r}'}/{r'^2}$, and extends all the way to infinity.   As shown in reference \cite{XLX:SJAM:2013}, the line charge density  $\lambda(t)$ \footnote{Note that in reference \cite{XLX:SJAM:2013}, the line charge density was denoted as $\rho(t)$.  } is related to the Mellin transform of $M_n(u)$ (as a function of $n$), and is an oscillatory function of $t$.  

To further reduce the computational cost, we truncate the line integral and discretize the line image using Gaussian quadrature.  $\chi(\rv, \rv')$ can then be expressed by $M$  point-like image charges and a few correction terms:
\begin{equation}
\chi(\rv,\rv') = \sum\limits_{m=0}^{M}
\frac{q_m}{4\pi\epsilon|\rv-\rv_m|}
+ \sum_l \chi_l P_l(\cos\theta)
\label{image-formula}
\end{equation}
where $M$ image point charges are located at $\rv_m$, with magnitude $q_m$. The correction terms are due to the truncation of the line integral. Numerical tests showed that $4$ images and $L=1$ corrections provide result with error less than $1\%$ in computing the self-energy of a charge. All relevant details of $q_m$, $\mathbf{r}_m$ and $\chi_l$ can be found in reference \cite{XLX:SJAM:2013}.  For not too small $r$, Eq.~(\ref{image-formula}) is much more efficient than the Kirkwood expansion,  Eq.~(\ref{Green_rf_series}).  

If the source point approach the center of the sphere, $r \ll R$, the Kirkwood expansion converges rapidly. We can use leading term of the Kirkwood expansion to approximate the Green's function:
\begin{equation}
\chi(\rv,\rv' = 0) = - \frac{\kappa}{4\pi\epsilon(1 + \kappa R)}.
\end{equation}

\subsection{Grand canonical Monte Carlo}
\label{sec:GCMC}
The system parameters used in our multi-scale simulation are as follows. The radius of simulation cavity is fixed to be $100\AA$.  The total ion number (including both species ) varies from 15 to 40.  The corresponding range of Debye length is between $24 - 50 \AA$.  Hence the system size is not much larger than the Debye length, and it is a nontrivial matter to cancel the influences of boundary and restore the physics of an infinite system.  All ions have hardcore diameter $d = 7.5\AA$.  

We simulate the subsystem ${\mathcal A}$ with an effective Hamiltonian Eq.~(\ref{H_A-eff}), with the reaction potential computed using Eq.~(\ref{image-formula}).  As emphasized in the preceding section, we must use grand canonical Monte Carlo to simulate this system. Use of canonical Monte Carlo would suppress fluctuations of total charge inside the cavity and therefore leads to substantial errors.  We will demonstrate this point in great detail below.

%\subsection{Grand Canonical simulation}
%\begin{comment}
The probability density function of a microstate with $N$ ions $\{q_1,\ldots,q_N\}$ at  ${\mathbf X}=\{\xv_1,\ldots, \xv_N\}$ is given by the grand canonical Gibbs distribution:
\begin{equation}
p_{N}(\mathbf{X})\prod_i^N d^3\xv_i = \frac{1}{\Xi}
\left[ \prod \limits_{s}^N\frac{e^{\beta\mu_sN_s}}
{N_s!\Lambda_s^{3N_s}}\right]
e^{-\beta H^{\mathcal A}_{\rm eff}}
\prod_i^N d^3\xv_i.
\end{equation}
Note that the grand canonical partition function $\Xi$ is dimensionless, whilst the dimension of $p (\mathbf{X})$ is $L^{-3 N}$.  To perform numerical simulation, however, it is mandatory to deal with discrete probabilities.  Therefore we discretize the simulation domain, with elementary length unit $|\delta \xv| $.  The probability of a discrete microstate is then
\begin{equation}
\pi_{N}(\mathbf{X})=\frac{1}{\Xi}
\left[ \prod \limits_{s}^N \frac{e^{\beta \mu_s N_s}
|\delta \xv|^{3N_s}}{N_s! \Lambda_s^{3N_s}}\right]
e^{-\beta H^{\mathcal A}_{\rm eff}}.
\label{pi-EQ}
\end{equation}
% where $|\delta \xv|$ is the distance between neighboring discretized positions of ions.

The Grand Canonical Monte Carlo simulation consists of three steps: {\em displacement, insertion} and {\em removal} \cite{Valleau-1980}:

\vspace{3mm}

\noindent {\em (1) Displacement.}\quad We select an ion randomly, and let its charge and position be $q_i$ and $\xv_i$.  The part of total energy of the system that depends on $q_i$ is  given by
\ba
E_i &=& \sum \limits_{j\neq i}
\Big[ q_i q_j G(\xv_i, \xv_j)  + w(\xv_i -\xv_j) \Big]
+  \frac{q_i^2}{2}\chi(\xv_i, \xv_i ) . \nonumber\\
%&+& %q_i\Phi_{\rm cav}
\label{E_i-general}
%\mbox{hardcore repulsion.} \nonumber
\ea
Then we choose a destination $\xv_i'$ for $q_i$ according to a flat probability density function defined in a cubic cell with a given size and centered at $\xv_i$, and let the new energy be $E_i'$. The change of the total energy of the system is
\ba
\Delta E &=& E'_i - E_i.
\label{Energy_dis}
\ea
The displacement is accepted with probability
\be
{\rm acc} \left( {\bf X}
\rightarrow {\bf X'} \right)
= \min \left\{ 1,e^{ - \beta\Delta E} \right\}
\ee
Evidently, if any hardcore constraint is violated in the destination state, $\Delta E = \infty$, and the displacement is rejected with probability one.  If the center of an ion across the boundary of simulation domain in an attempted move, it is also rejected.  In another word, the boundary behaves as a hard wall to the center-of-mass of all ions.  It is easy to see that this transfer probability satisfies the detailed balance with respect to the equilibrium Gibbs distribution Eq.~(\ref{pi-EQ}):
\be
\pi({\bf X}) \,  {\rm acc} \left( {\bf X}
\rightarrow {\bf X'} \right)
= \pi({\bf X'})  \, {\rm acc} \left( {\bf X'}
\rightarrow {\bf X} \right).
\ee

\vspace{3mm}

\noindent{\em (2) Insertion.} \quad To mimic the fluctuations of particle numbers in an open system, we need to insert and remove particles.  In a micro state with $N_s$ particles of species  $s$, we first  randomly select a position $\xv$ in the cavity according to a uniform probability distribution. The latter can be generated by introducing three independent random variables $\eta_1$, $\eta_2$, $\eta_3$ that are uniformly distributed in the interval $[0,1]$, and express the spherical coordinates of the insertion point $\xv$ as
\ba
r =  \eta_1^{1/3} R,  \quad
\theta = \cos^{-1}(1-2 \eta_2),
\quad \phi = 2\pi \, \eta_3.
\ea
It is straightforward to verify that the Jacobian $\partial(x,y,z)/\partial(\eta_1,\eta_2,\eta_3)$ is a constant independent of $\eta_1$, $\eta_2$, $\eta_3$, where $x,y,z$ are the Cartesian coordinates.  We now choose a species $s$ randomly and insert an ion $q_s$ at $\xv$ with probability\begin{equation}
{\rm acc} \left( N_s \rightarrow  N_s+1 \right)
 = \min \Bigg \{1,\frac{ e^{\beta( \mu_s^* -  E_i)   } }
 {N_s+1} \Bigg\},
\end{equation}
where $E_i$ is the energy change due to the insertion operation, given by Eq.~(\ref{E_i-general}), whilst $\mu_s^*$ are the effective chemical potential of the $s$th species , given by
\begin{equation}
\mu_s^*=\mu_s+k_{\rm B} T \log \frac{V} {\Lambda_s^3}.
\end{equation}
%\xing{Need explanation for $\mu_s^*$ here?}

\vspace{3mm}

\noindent {\em (3) Removal.} \quad In a micro state with $N_s$ particles of species  $s$, we randomly choose a particle with (a randomly chosen) species  $s$, and remove it with probability
\begin{equation}
{\rm acc} \left(N_s \rightarrow N_s - 1 \right)
= \min \left\{1,{ N_s \, e^{-\beta( \mu_s^* -  E_i) }}\right\}
\end{equation}
where $- E_i$ is the energy change of the removal operation, with $E_i$ again given by Eq.~(\ref{E_i-general}).

It is straightforward to verify that the probabilities of insertion and removal satisfy detailed balance with respect to the equilibrium grand canonical distribution Gibbs distribution Eq.~(\ref{pi-EQ}):
\ba
&& \pi_N \left( \frac{|\delta \xv |^3}{V}\right)
{\rm acc} \left(N_s  \rightarrow N_s +1 \right)
\nonumber\\
&=&
 \pi_{N+1} \left( \frac{1}{N_s +1} \right)
 {\rm acc} \left(N_s +1 \rightarrow N_s  \right).
\ea
The factor ${|\delta \xv|^3}/{V}$ in the LHS is the probability of choosing one particular point inside the (already discretized) simulation domain, whereas the factor $1/(N_s+1)$ in the RHS is the probability of choosing a particular particle with species  $s$.

% Note that for each species , the frequency of insertion and removal procedures should be equivalent.
%\end{itemize}

\subsection{Comparison with Other Simuation Methods}
\label{sec:comparison}

Three other simulation methods are used to compare with our multi-scale reaction potential GCMC simulation method (RP+GCMC).  To demonstrate the artifacts due to periodic boundary conditions, we  use the popular Ewald summation GCMC method (PBC+GCMC) to conduct a small scale simulation.  To demonstrate the artifacts due to suppression of charge fluctuations, we apply a small scale simulation on reaction potential model using canonical Monte Carlo (RP+CMC).  Finally, we conduct a large scale simulations with system sizes at least ten Debye lengths to obtain standard results (STD), with which all other simulations are compared.  To make the comparison meaningful, we adjust parameters such that the ion densities in the bulk are equal in all simulations.  Furthermore, all small scale simulations have equal size of simulation domains.  

%We also measure the effective chemical potential for each ion species  using the Widom method \cite{Article:6}. We insert an ion of species $s$ at a random position, and sample the potential acting on it due to all other ions.  The effective chemical potential of species  $s$, $\mu_s$, is then given by
%\ba \mu_s^*&=&\mu_s+k_{\! B} T \log\frac{V}{\Lambda^3}\nonumber\\
%&=&k_BT \log(N_s+1) - k_BT \,  \log \, \langle e^{-\beta \Psi} \rangle.
%\label{mu-Widom} \ea
%The computed chemical potential is then used to conduct various grand canonical MC simulations.

\vspace{3mm}
\noindent{\em (1) Grand canonical Ewald summation MC with Periodic Boundary Conditions 
(PBC+GCMC)}.  \quad
We choose cubic simulation domain with volume equal to $4 \pi R^3/3$. To avoid divergence when summing over periodic image charges, the total charge in the simulation box must vanish in every micro state.  Furthermore, one still need to specify some ``boundary conditions'' at infinity.  We shall use the popular ``conductor boundary condition'', where the average potential inside the simulation box vanishes, and there is no dipolar term in the total free energy.  For details, see the textbook by Frenkel and Smit \cite{frenkel2002book}.  This choice is consistent with the condition Eq.~(\ref{Gauge1}) that we use in the analysis of LPB with PBC. 

The total ion number however can fluctuate according to the standard grand canonical ensemble theory.  In accordance with these constraints, ions are inserted and deleted at random as pairs \cite{Valleau-1980}.  

\vspace{3mm}
\noindent{\em (2) Multi-scale Reaction Potential Canonical MC simulation (RP + CMC).} \quad
%\vspace{1mm}
This simulation is similar to RP+GCMC, with the only difference that the ion number of each species  is kept constant. 

\vspace{3mm}
\noindent {\em (3) Large Scale Canonical MC simulation with Hard Wall Boundary Conditions (STD).}
%\vspace{1mm}
\quad
We conduct a large scale canonical simulation using a spherical simulation domain with radius at least ten times of Debye length, and with hard wall boundary condition.  To void the artifacts due to boundary effects, only ions that are more than five Debye lengths away from the boundary are used for data collection.  The simulated system typically contains about 2500 to 5000 ions.  To speed up the computation of total electrostatic energy, we use the recently developed  oct-tree algorithm \cite{Gan2014scm}.  We calculate various physical quantities via this method and use them as {\em standard results} for all later comparisons.

% \xing{\xing{Need system parameters here: size of simulation domain(\Yihao{STD:$R=500\AA$,RF:$R=100\AA$, Ewald:$L=161.1992\AA$}.The STD simulation contains 2500 particles,whereas other simulations contain 20 particles.), diameter of ion:\Yihao{$a=7.5\AA$}, Debye length(\Yihao{$\kappa_R=0.02064029\AA^{-1}$}).} }

%Two issues to be discussed: 1) artificial periodic image charges; 2) artificial constraint of charge neutrality.  Both introduce artifacts in the simulation. It appears that these two issues are actually related to each other.  In particular, by including all periodic image charges, we automatically get charge neutrality inside the simulation box!

\section{Results and Discussions}
\label{sec:results}
We compute various physical quantities using three small scale simulations, and compare them with the standard results (STD).  Disagreement with STD indicates the simulation strategy is incapable of capturing the correct physics of infinite systems.  

% To make the comparison meaningful, we adjust various parameters such that the mean particle numbers are identical in  all simulations ( except STD, of course).  

\begin{figure}[t!]
\includegraphics[width=0.37\textwidth] {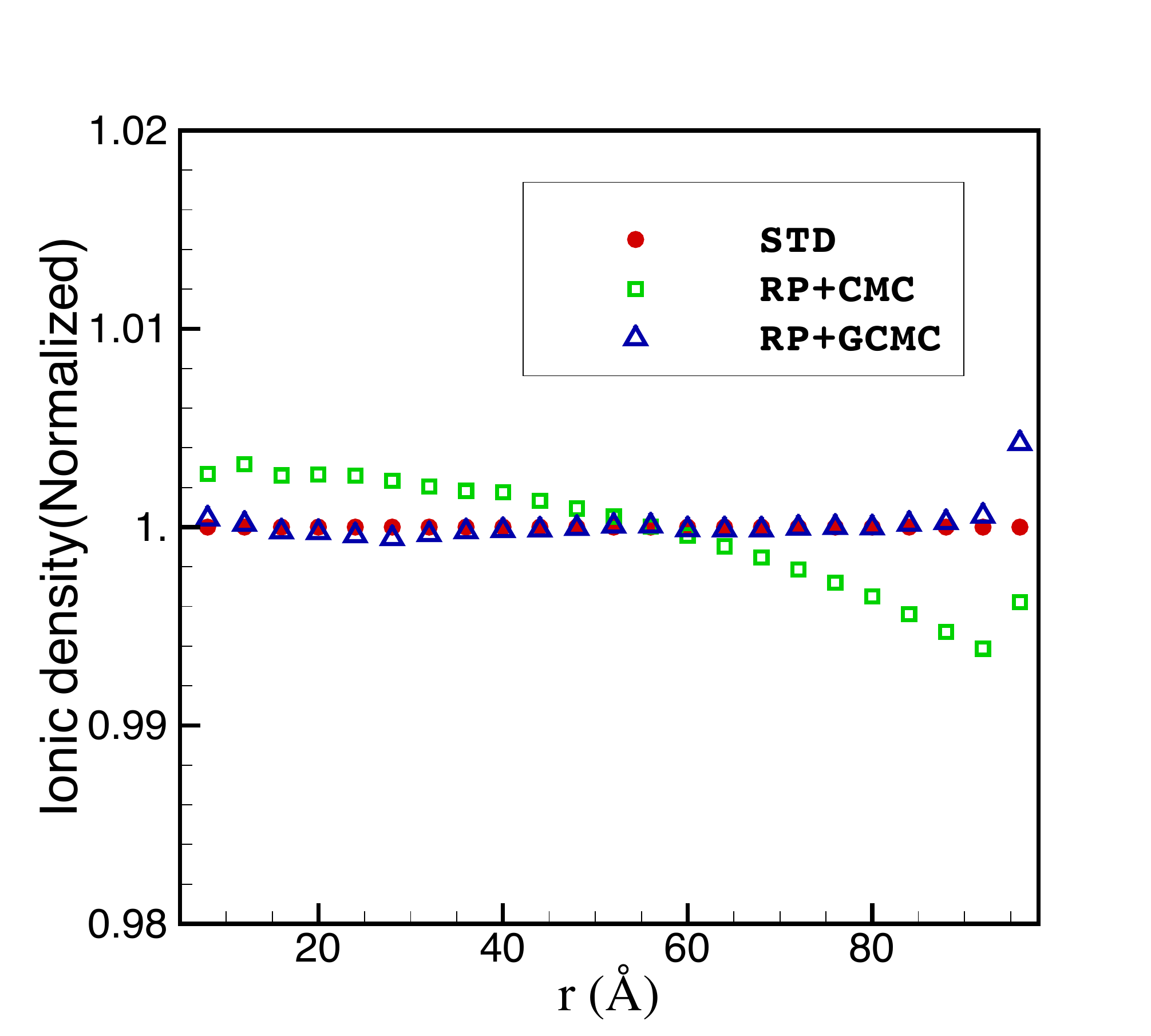}
\label{Fig:ion_density}
\caption{Ion density profile inside the simulation cavity, as a function of radius.  The density profile of RP+GCMC is almost flat inside the cavity, except in a boundary layer of approximately one ion diameter thick.  By contrast, The profile of RP+CMC  shows a much larger variation inside the cavity.  This is caused by the artificial constraint of charge neutrality inside the cavity.  The standard results are obtained using a large scale simulation, and therefore are completely flat in the plotted region. }
% ( \xing{Add the corresponding results without reaction potential correction.} }
\label{Fig:ion_density}
\end{figure}

\vspace{3mm}
\noindent {\bf (1) Average ion densities inside cavity} \quad
\vspace{1mm}

We measure the average ion density in the simulation cavity as a function of the radial distance, shown in Fig.~\ref{Fig:ion_density}.  If the multi-scale modeling is faithful, the resulting average ion density must be independent of radius. In another word, non-uniformity of ion density indicates that the artifacts of cavity boundary have not been properly cancelled.  In Fig.~\ref{Fig:ion_density}, we see that RP+GCMC simulation yields a much flatter density profile than RP+CMC.  For the latter case, there is a systematic tendency that the ion density is higher near the center than near the boundary, with an overall variation of about one percent. This is caused by the artificial constraint of charge neutrality inside the simulation cavity.  %\xing{In contrast to the GCMC simulation, in CMC simulation, the simulation cavity carries an excess charge that is opposite to the source charge, which attracts the source charge toward the center.  }

Note that both RP+GCMC and RP+CMC show a thin boundary layer (with thickness comparable with ion diameter) where the ion densities vary rather abruptly.  This is due to the short scale depletion effects of the hard wall boundary, and shows up actually in all simulations with hard wall boundary conditions.  Even though correction of these depletion effects are straightforward, it is more convenient just to exclude the boundary layer completely for data collection.    

%We do not plot the corresponding result for PBC+ Ewald summation method, because there the ion density is always flat, since PBC always restores the translational symmetry.

This comparison clearly demonstrates that we should use grand canonical, instead of canonical, ensemble when implementing the multi-scale simulation strategy.

\vspace{3mm}
\noindent {\bf (2) Correlation potential}\quad
\vspace{1mm}

\begin{figure}[t!]
\includegraphics[width=0.37\textwidth]{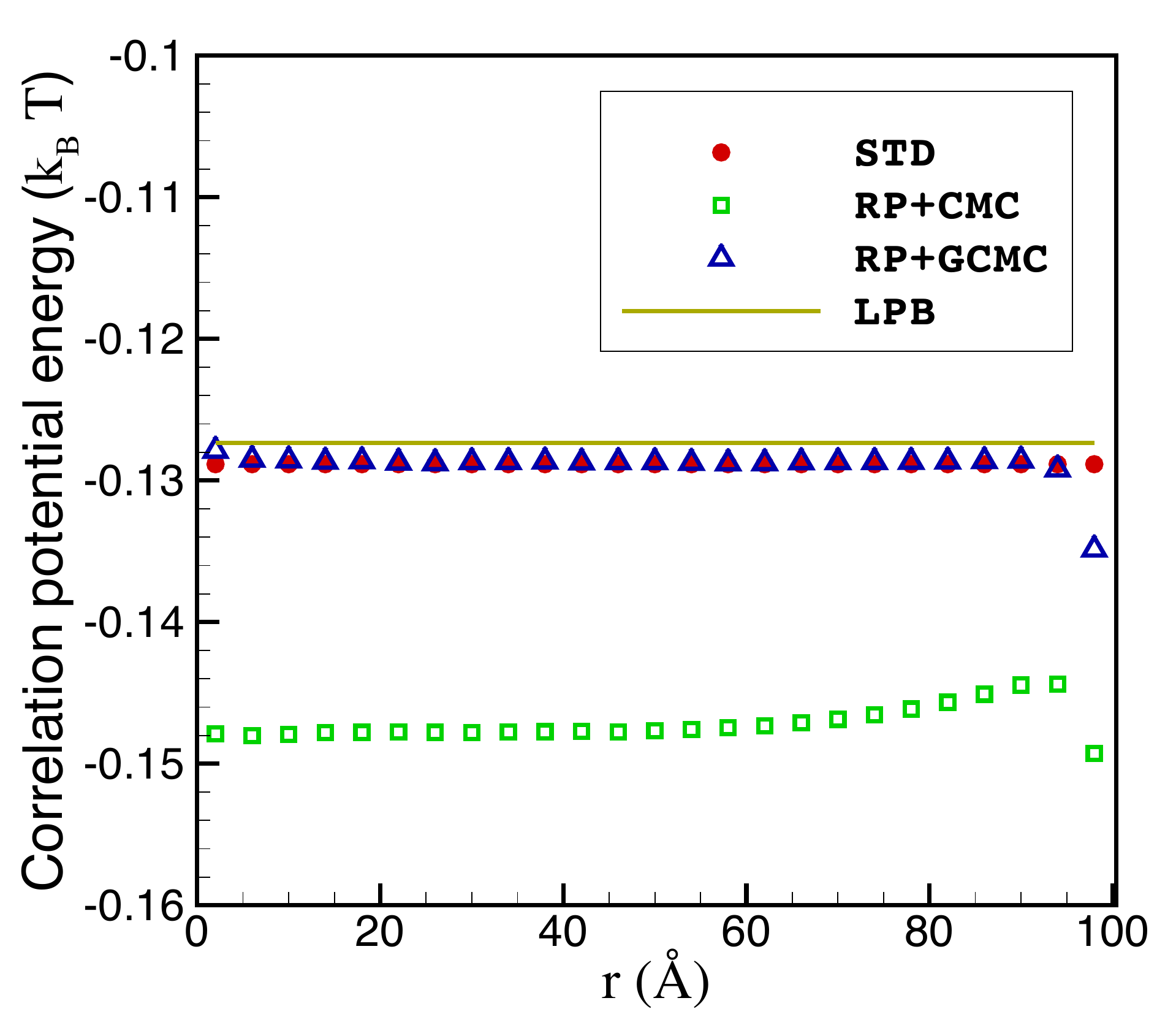}
\caption{Correlation potential acting on a positive ion, as a function of its position.  The classical Debye-H\"uckel theory works well in the low density regime.  The substantial errors in RP+CMC arise to the artificial constraint of charge neutrality inside the simulation cavity.  The small difference between DH and standard simulation results is caused by charge correlation effects. }
\label{Fig:correlation_potential}
\end{figure}

We measure the mean potential acting on an ion due to all other ions, which is usually called the {\em correlation potential}.  Because we are in the low density regime, the Debye-H\"uckel theory is applicable.  Indeed, as shown in Fig.~\ref{Fig:correlation_potential}, both RP+GCMC and STD agree well with prediction of DH.  By contrast, RP+CMC shows substantial deviation from the other results.  This is again due to artificial constraint of charge neutrality inside the cavity.  

This correlation potential is {\em not} measurable in PBC+GCMC simulation, because the summation over potential due to all periodic images is divergent.

\vspace{3mm}
\noindent {\bf (3) Screening charge density around a test ion} \\
%\vspace{1mm}

We fix an ion inside the simulation cavity and measure the screening charge density around, as a function of the distance to the fixed ion. This charge density can be obtained using data of pair correlation functions inside the whole simulation cavity.  The results have already been presented in Fig~\ref{Fig_screening-rho}.  Again, RP+GCMC agrees with STD up to high precision, whilst  both PBC+GCMC and RP+CMC yield results that are substantially different.  

Theoretically, LPB predicts that the charge density has the form of screened Coulomb:
\be
\rho_q(r) = - \epsilon \nabla^2 \phi(\rv)
= - \frac{q\, \kappa^2 e^{-\kappa r}}{4 \pi r}
+ q \, \delta(\rv).
\label{screening-charge}
\ee
Hence if we plot $r \rho_q(r)$ v.s. $r$ in log scale, we should obtain a straight-line, with slope given by the inverse Debye length.  As shown in Fig~\ref{Fig_screening-rho}, this agrees remarkably well with the simulation results using RP+GCMC.  

% Therefore, use of PBC or RP+CMC would be erroneous, because one would not even be able to extract the correct Debye length from simulation data. 

%It seem a little coincidental to us that the results of RP+CMC and PBC+Ewald are quite similar to each other in this case. 

We also plot the the analytic result of LPB with PBC, Eq.~(\ref{rho_q-PBC}) in Fig~\ref{Fig_screening-rho} as red diamonds (PBC+LPB).  This agrees very well with Ewald method in the far field.  Since Eq.~(\ref{rho_q-PBC}) differs from Eq.~(\ref{screening-charge}) only by the periodic images and their screening cloud, the latter can be identified a main source of errors in simulations using periodic boundary conditions.    

\begin{figure}[tb!]
\includegraphics[width=0.37\textwidth]{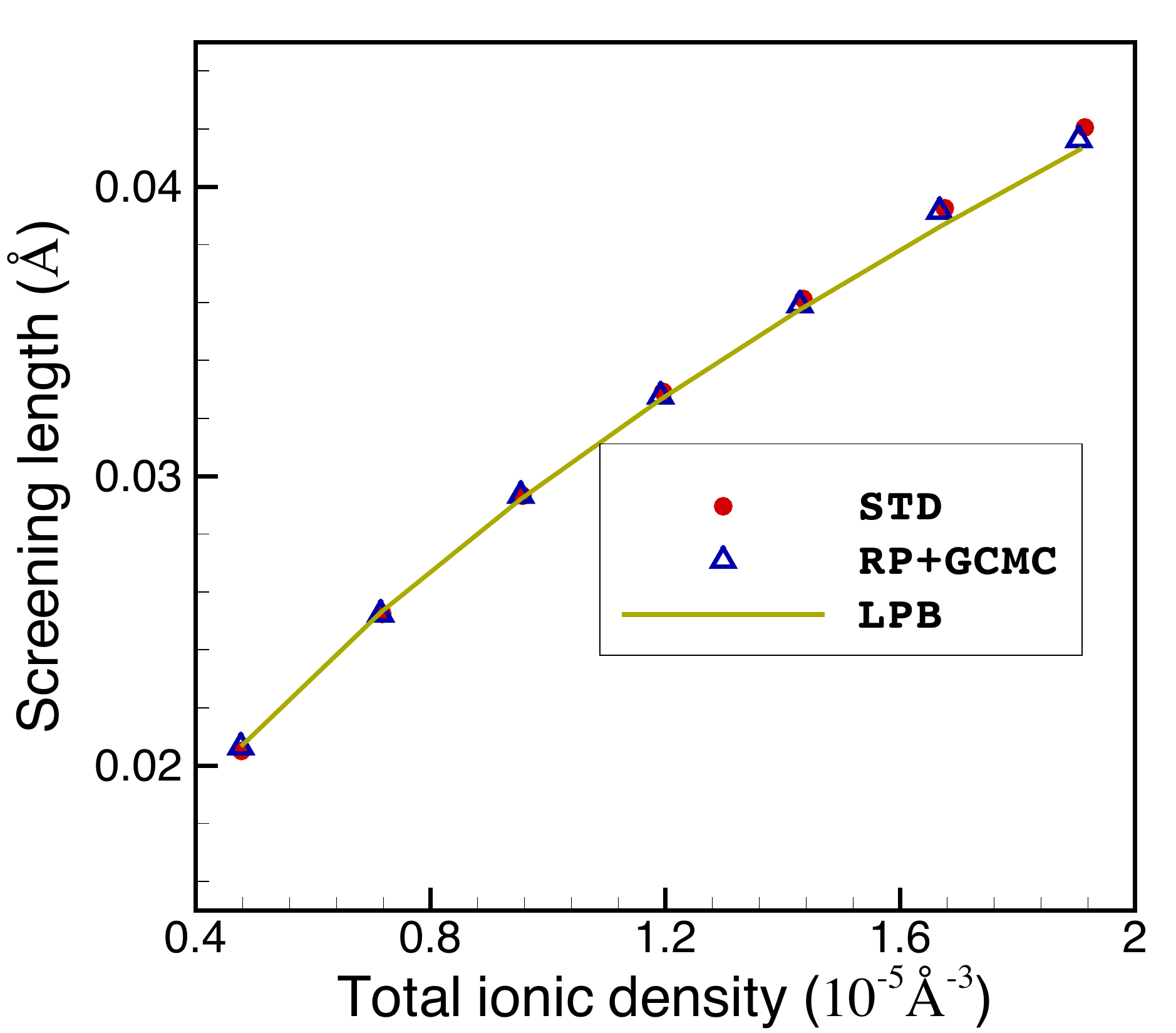}
\caption{The screening length as determined from large scale simulation and from multi-scale GCMC simulation.  The solid curve is the prediction of PB theory.  }
\label{fig:decay_length}
\vspace{-3mm}
\end{figure}

The charge densities obtained using both RP+GCMC and STD can be used to determine the screening length as a function of ion density.  The results are shown in Fig.~\ref{fig:decay_length}.  Also shown there is the prediction of screening length by LPB, Eq.~(\ref{kappa-PB}), which agrees rather well with simulation results.  This shows that  LPB is indeed applicable for the ion densities studied in this work.

% \xing{Alternatively, in Eq.~(\ref{Fig_illustration:d}) we plot the ion density around an ion fixed at a positive that is off the center. }

\vspace{3mm}
\noindent  {\bf (4) Internal energy per ion} \quad
\vspace{1mm}

% \xing{here we need to discuss in detail how the internal energy is calculated in each simulation. }

%\xing{*****************************************}

We use different methods to compute the internal energy of the system per ion.  The results are  shown in Fig~\ref{Fig_internal-energy-1}.   The total internal energy of the system can be easily calculated in terms of the correlation potential acting on each ion:
$E = \frac{1}{2} \sum_{i} q_i \psi_i$. 
Again our multi-scale GCMC simulation yields the same results as the large scale simulation.   Both multi-scale CMC and PBC+GCMC yield incorrect results for the internal energy.  In the same figure, the solid line (DH) is the prediction of {Debye-H\"uckel theory} in open space (the first term in RHS of Eq.~(\ref{chi-hardsphere})), which agrees with STD and RP+GCMC very well.  
The dashed line is the prediction of {Debye-H\"uckel theory} with PBC as well as leading order correction of hardcore taken into account (Eq.~(\ref{chi-hardsphere})).  It appears that Eq.~(\ref{chi-hardsphere}) takes into account most of the artifacts in PBC+GCMC simulation if the density is sufficiently low.  As the density increases, Eq.~(\ref{chi-hardsphere}) deviates from PBC+GCMC more significantly.  

% In particular, PBC+Ewald substantially overestimate the correlation energy.  }

% \xing{Need to specify how to compute the internal energy using multi-scale simulation The internal energy of an ion at radius $r$ is calculated as follows. }

%\begin{figure}
%\includegraphics[width=0.37\textwidth]{chemical_potential.eps}
%\caption{The chemical potential, computed using three methods: 1) large scale simulations (STD) and 2) mutli-scale GCMC (RF+GCMC), and 3) grand canonical Ewald summation. It appears that all these methods  yield essentially the same results.  \xing{delete this figure!}}
%\label{Fig_TOT}
%\caption{How is screening length defined here? }
%\end{figure}

\vspace{3mm}
\noindent {\bf (5) Total charge inside the simulation cavity} \quad
\vspace{1mm}

\begin{figure}[t!]
\includegraphics[width=0.35\textwidth]{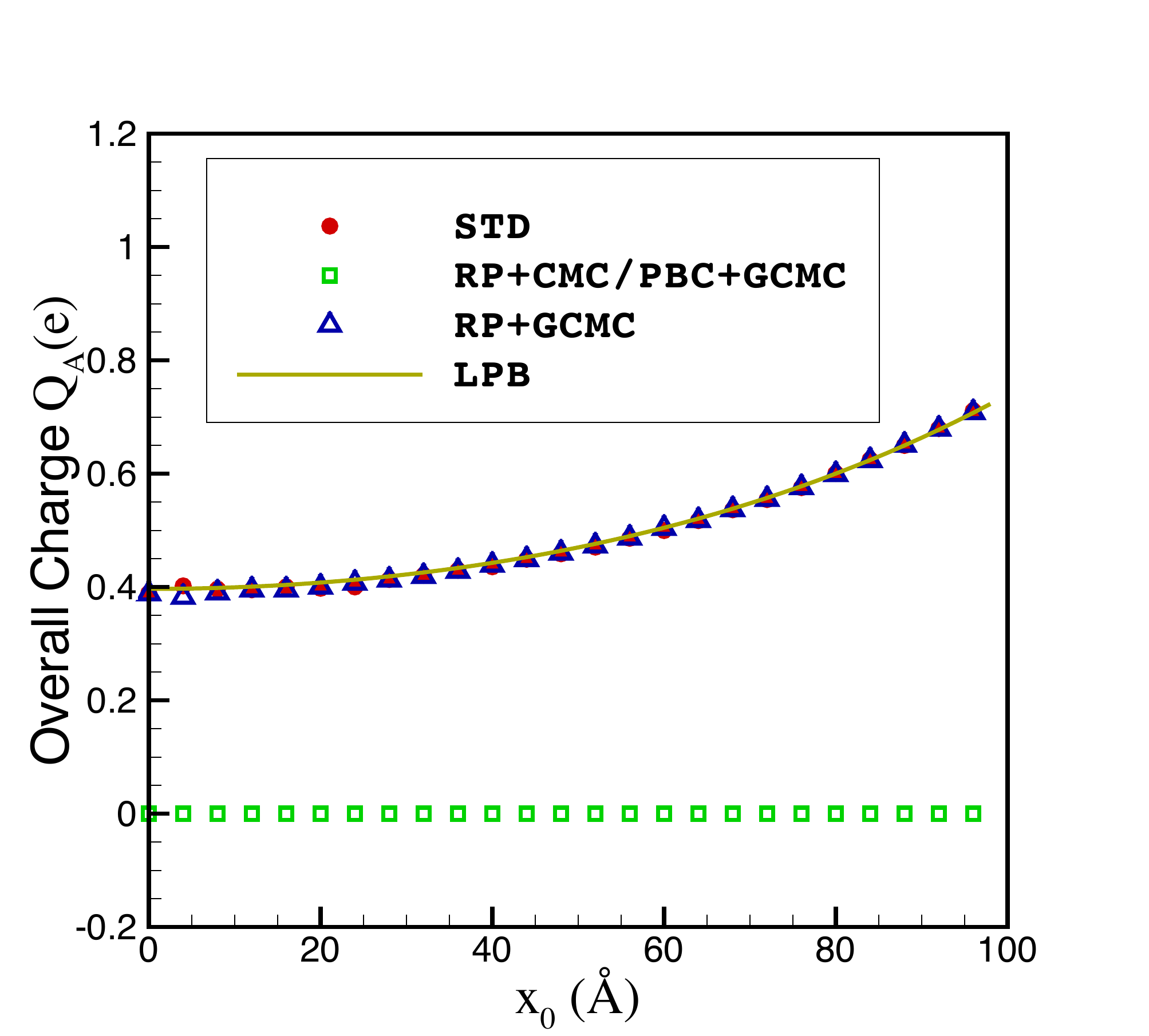}
\label{Fig_Totc:a}
\caption{ A test ion is fixed at $r$ inside cavity, and the total mean charge in the cavity is plotted as a function of $r$.  Both STD and RP+GCMC agree well with LPB, Eq.~(\ref{Q-tot-cavity}). } 
\vspace{-3mm}
\end{figure}
 
As a stringent test of the faithfulness of our multi-scale method, we fix a positive ion at a distance $r$ from the origin, and measure the total charge inside the cavity.  The geometry is illustrated in Fig.~\ref{Fig_illustration}.  The total mean charge inside the cavity is nonzero, because screening can not be perfect in a finite volume.  In the framework of LPB, the total charge density is given by Eq.~(\ref{screening-charge}). Hence the total charge inside the cavity can be easily calculated:
\ba
\Delta Q (r) &=& q - \frac{q \kappa^2}{4 \pi}
\int_{|\xv| \leq R} \frac{e^{-\kappa |\xv - \rv|}} {|\xv - \rv|} d^3 \xv
\nonumber\\
&=& \frac{q\, e^{- \kappa R} (\kappa R+1)
 \sinh (\kappa r)}{\kappa r}.
\label{Q-tot-cavity}
\ea
This result, shown as the solid solid curve in Fig.~\ref{Fig_Totc:a}, agrees remarkably well with STD and RP+GCMC.  Again, this indicates that 1) LPB works well in the density regime under study; 2)  our RP+GCMC simulation method faithfully captures the physics of an infinite electrolyte.  By contrast, for RP+GCMC, this total charge is identically zero, due to the condition of charge neutrality.

\begin{figure*}[tbh!]
\subfigure[]{
\includegraphics[width=0.31\textwidth]{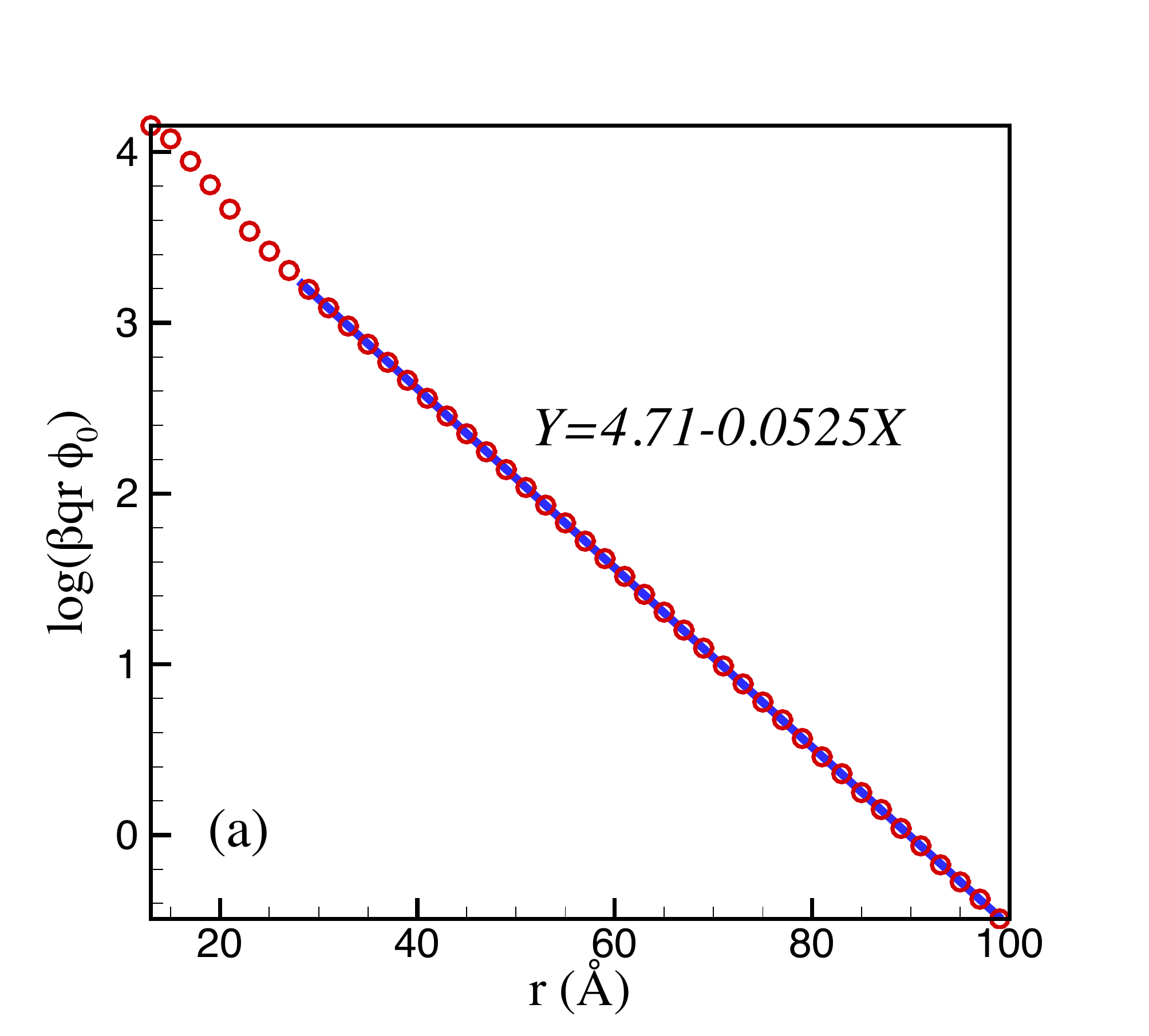}
\label{Dec_Colloid:a}
}
\subfigure[]{
\includegraphics[width=0.32\textwidth]{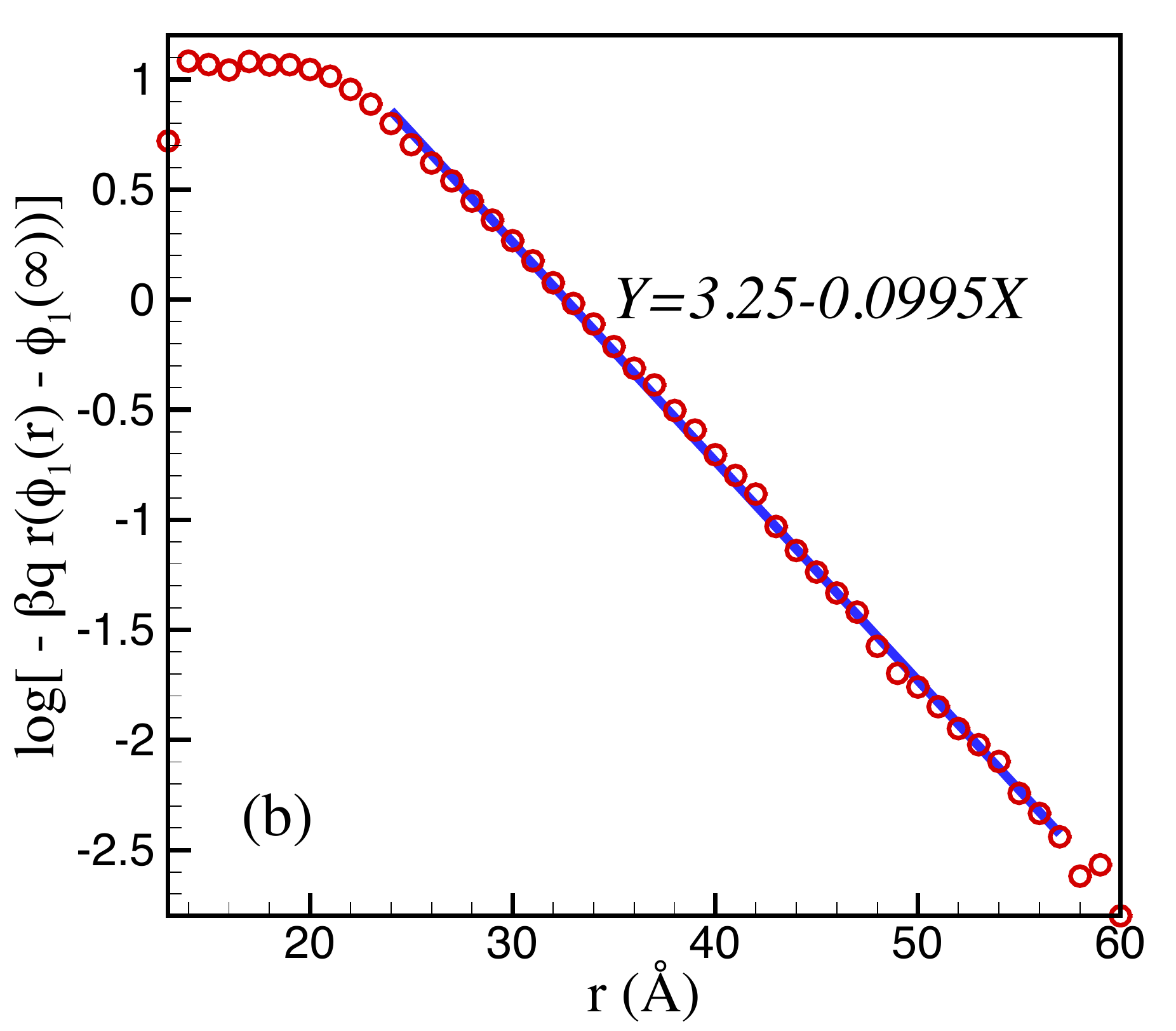}
\label{Dec_Colloid:b}
}
\subfigure[]{
\includegraphics[width=0.31\textwidth]{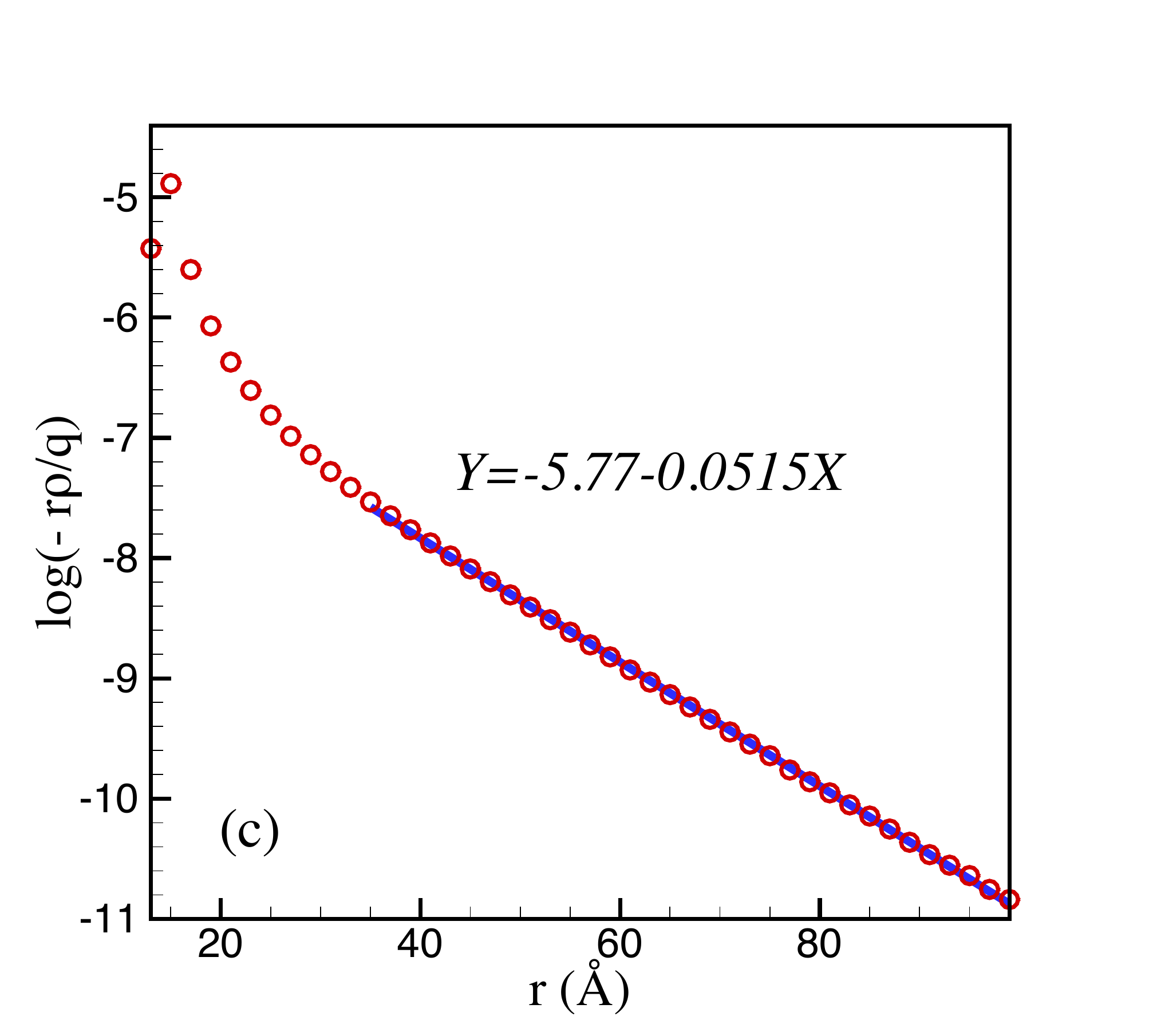}
\label{Dec_Colloid:c}
}
\caption{ Potential and ion density profile near a charged spherical colloid.  (a) $\phi_0(r)$, and (b) $\phi_1(r) - \phi_1(\infty)$.  These potentials are defined in Eq.~(\ref{phi-til-expansion}).   (c) Net charge density around the charged colloid. The decay length scales of (a) and (c) are approximately the same, and are about twice of that of (b), in agreement with theoretical predictions.  }
\label{Dec_Colloid}
\end{figure*}

\vspace{3mm}
\noindent {\bf (6) Ion densities near a charged colloid} \quad
\vspace{1mm}

Finally to test the applicability of our multi-scale simulation strategy for inhomogeneous systems, we insert a uniformly charged colloid with charge $Q_0 = 20 e$ and radius $R_0 = 10 \AA$ at the center of the simulation cavity, and measure the local potential $\tilde{\phi}(\rv,q)$ acting on an ion fixed at $r > R_0$, due to both the charged colloid as well as all other ions.   We set the dielectric constant of the colloid to be the same as that of the solvent, so that there is no image charge effect.  Note that $\tilde{\phi}(\rv,q)$ also depends on the ion diameter $d$.  We do not show this dependence in order to avoid cluttered notations. 

Note that the potential $\tilde{\phi}(\rv,q)$  is different from the average potential $\phi(\rv)$ in the absence of the test ion.  We shall call the former {\em the conditional mean potential}, and the latter {\em the unconditional mean potential}.  Both quantities can be directly measured using simulation data. The difference between them arises because the test ion influences the distribution of other ions.  The difference between $\tilde{\phi}(\rv,q)$ and $\phi(\rv)$ encodes essential information about ionic correlations in this inhomogeneous system.  We shall numerically compute $\tilde{\phi}(\rv,q)$ and clarify its relation with $\phi(\rv)$. This demonstrates that our multi-scale reaction potential GCMC is capable of capturing the ionic correlations faithfully in an inhomogeneous system.  

% Even though at the ion densities being simulated in this work, ionic correlations are negligibly small, they can nevertheless be extracted and analyzed. If the influences of domain boundary is not completely cancelled in the bulk, they will be visible at this order, since the domain boundary is charge-neutral.

The conditional mean potential $\tilde{\phi}(r, q)$ can be expanded in terms of the charge $q$:
\ba
\tilde{\phi}(r,q) =  \phi_0 (r) + q \, \phi_1(r) + O(q^2).
\label{phi-til-expansion}
\ea
It is important to realize that $\phi_0(r)$ is not the same as the unconditional mean potential $\phi(r)$.  Their difference is caused by the hardcore of the test ion.  Nevertheless, in the far field regime where LPB is applicable, we can show that $\phi_0(r)$ and $\phi(r)$ are proportional to each other, and both assume the form of screened Coulomb potential:
\be
\phi_0 (r), \phi (r) \propto \frac{e^{-\kappa r} }{r}.
\ee
The potential $\phi_1 (r)$ is not completely known.  In reference \cite{Lu:2014uq}, it was analytically calculated near a uniformly charged {\em flat surface}, by perturbing around the nonlinear PB theory.   It was found that $\phi_1 (r)$ has the form 
\be
\phi_1 (r) = \phi_1(\infty)
+ f(r) \times \frac{e^{-2 \kappa r}}{r},
\label{Phi_1-slow}
\ee
where $f(r)$ is a slowly varying function (comparing with exponential), and $\phi_1(\infty)$ is the reaction potential in the bulk. This result is expected to be valid in the present spherical geometry.    

Now using Eq.~(\ref{phi-til-expansion}) we can express $\phi_0(r), \phi_1(r)$ as linear combinations of $  \tilde{\phi}(r, +q)$ and $  \tilde{\phi}(r, -q)$:
\ba
\phi_0 (r) &=& \frac{1}{2} \Big[   \tilde{\phi}(r, +q) + \tilde{\phi}(r, -q) \Big],\\
\phi_1 (r) &=&  \frac{1}{2q} \Big[  \tilde{\phi}(r, +q) - \tilde{\phi}(r, -q) \Big].
\ea
Hence $\phi_0(r), \phi_1(r)$ can be computed directly using simulation data.  In Fig.~\ref{Dec_Colloid:a} and \ref{Dec_Colloid:b}, we plot $r\, \phi_0(r)$ and $r\, [ \phi_1(r) - \phi_1(\infty) ]$  as a function of $r$ in log scale.  The far field asymptotics of both functions indeed behave as screened Coulomb, with their characteristic decay lengths differ by a factor of two.  This demonstrates that in our multi-scale GCMC simulation, the influences of the boundary of simulation cavity are properly cancelled, so that the simulation data can be used to analyze subtle ionic correlations.   Also the fact that $\log \left(r\, \phi_1 (r) \right)$ scales linearly with $r$ in the far field indicates that the {\em slow function} $\chi(r) $ is indeed changing very slowly.

%Also the fact that It also demonstrates that the influences of domain boundary has been {\em completely} cancelled, and our multi-scale simulation strategy is capable of simulation subtle correlation effects, even though the simulation domain is rather small.

The potential of mean force (PMF) of a test ion (with positive charge $q$) can be obtained from the local potential Eq.~(\ref{phi-til-expansion}) via the method of Debye charging:
\ba
U(r, q) &=& \int_0^q  \left[ \tilde{\phi}(r,q) - \tilde{\phi}(\infty,q) \right]
\, dq + U(r, 0).
\nonumber\\
&\approx&  q\, \phi_0 (r) + O(e^{-2\, \kappa r})
\ea
The constant of integration $U(r, 0)$ is the free energy cost of bringing a neutral particle from the bulk to $r$. In a symmetric electrolyte, $U(r, 0)$ decays twice as fast as $\phi_0 (r)$, and therefore makes no contribution to the leading order far field asymptotics of $U(r, q)$.  Now the average density $\rho_+(r)$ of positive ions is related to the PMF $U(r, q)$ via the following well known Gibbs distribution.  Hence the total charge density is 
\ba
\rho(r) = q \rho_0 \left[ e^{- \beta U(r, q) } -  e^{ \beta U(r, - q) } \right]
\approx - 2 q^2 \beta \, \phi_0 (r).
\ea
That is, the deviation of ion density from its bulk value also behaves as a screened Coulomb in the far field.  In Fig.~\ref{Dec_Colloid:c} we plot the corresponding simulation results and show that indeed $\rho(r)$ obeys the law of screened Coulomb, with the same characteristic decay length $1/\kappa$.

\section{Conclusion and Acknowledgement}
\label{sec:conclusion}
We have demonstrated the artifacts due to periodic boundary condition in the simulation of electrolytes.  These artifacts are caused by the periodic image charges and the constraint of charge neutrality inside simulation domain, both unphysical from the perspective of mimicking an infinite system.   One can always avoid these artifacts by simulating a much larger system and throwing away data from the region near the boundary.  This leads to waste of computational resource, but is frequently done in simulations.  Our multi-scale reaction potential GCMC provides a more efficient alternative.  Combining reaction potential modeling and grand canonical Monte Carlo, this method cancel most of the electrostatic artifacts introduced by the boundary of simulation data, and hence faithfully capture the physics of an infinite system even in a very small scale simulation (with linear size about three Debye lengths).   If we combine this multi-scale GCMC method with appropriate fast algorithm for evaluation of electrostatic energy (such as the oct-tree algorithm), we arrive at a method for large scale MC simulation of electrolytes, that will constitute a competitive alternative to the current methods based on molecular dynamics.   This will be explored in a future publication.  

We thank NSFC (Grants No. 11174196 and 91130012) for financial support, the High Performance Computation Centre ($\Pi$) at Shanghai Jiao Tong University for computation resources.  We also thank  Wei Cai for interesting discussions.  

%\bibliographystyle{elsart-num}
%\bibliographystyle{unsrt}
%\bibliography{merged_bib}

\end{document}